\documentclass[a4paper,11pt]{article}
\pdfoutput=1 

\usepackage{jcappub} 

\usepackage{amsmath}
\PassOptionsToPackage{hyphens}{url}
\usepackage{setspace}
\usepackage{caption}
\usepackage{subcaption}
\usepackage[T1]{fontenc}
\usepackage[dvipsnames]{xcolor}

\usepackage{tikz}
\usetikzlibrary{
    arrows.meta,      
    calc,             
    decorations.markings, 
    positioning       
}
\usetikzlibrary{decorations.pathmorphing}
\usetikzlibrary{patterns}
\usetikzlibrary{shadings}

\title{Long(itudinal) live the helioscope: novel bounds on solar dark photons from conversion on reflecting surfaces}

\author[]{Jonas Frerick}

\affiliation[]{Dipartimento di Fisica, Sapienza Università di Roma \& Sezione INFN Roma 1\\
                Piazzale Aldo Moro 5, 00185 Rome, Italy}

\emailAdd{jonas.frerick@uniroma1.it}

\abstract{
Conventional helioscope searches are insensitive to the longitudinal component of the solar dark photon (DP) flux, which dominates the solar emission for DP masses $m\lesssim 1\,$eV.
In this paper, we demonstrate that this limitation can be overcome: using the interface conditions that follow from Maxwell's equations in the presence of a dark photon, we show that a relativistic longitudinal DP flux impinging on a telescope mirror produces an outgoing photon flux. We apply this principle to X-ray helioscopes which employ grazing incidence mirrors. As our main result, we find that BabyIAXO would be competitive with the current world-leading constraints from XENON1T if it can keep sensitivity down to energies of $10\,$eV. Both IAXO and a proposed upgraded version would be able to reach a factor of $\sim 2.5$ and $\sim 13$ deeper than BabyIAXO, respectively, and thus probe a large fraction of unexplored parameter space for $10^{-5}\,\text{eV}\lesssim m\lesssim 0.75\,$eV.
}

\begin{document}

\maketitle
\flushbottom

\section{Introduction}
\label{introduction}

The continued search for physics beyond the Standard Model (SM) has led to a large variety of searches for new particles. Alongside prominent candidates such as WIMPs \cite{Arcadi:2017kky} and axions \cite{DiLuzio:2020wdo}, the dark photon (DP) is a particularly well-motivated example \cite{Okun:1982xi, Holdom:1985ag, Fabbrichesi:2020wbt}. The DP is a new vector boson that couples to the SM predominantly through kinetic mixing with the photon. This distinguishes it from a generic $Z'$ boson, which may couple directly to SM currents or mix with the $Z$ boson \cite{Langacker:2008yv, Bauer:2018onh, Frerick:2023xnf}. The resulting phenomenology has motivated a broad programme of experimental and theoretical studies of the DP parameter space \cite{ Caputo:2021eaa, Essig:2013lka, Caputo:2025avc, Caputo:2026pdw}.

\begin{figure}[ht!]
    \centering
    \includegraphics[width=0.9\linewidth, trim = 2cm 0 0 0]{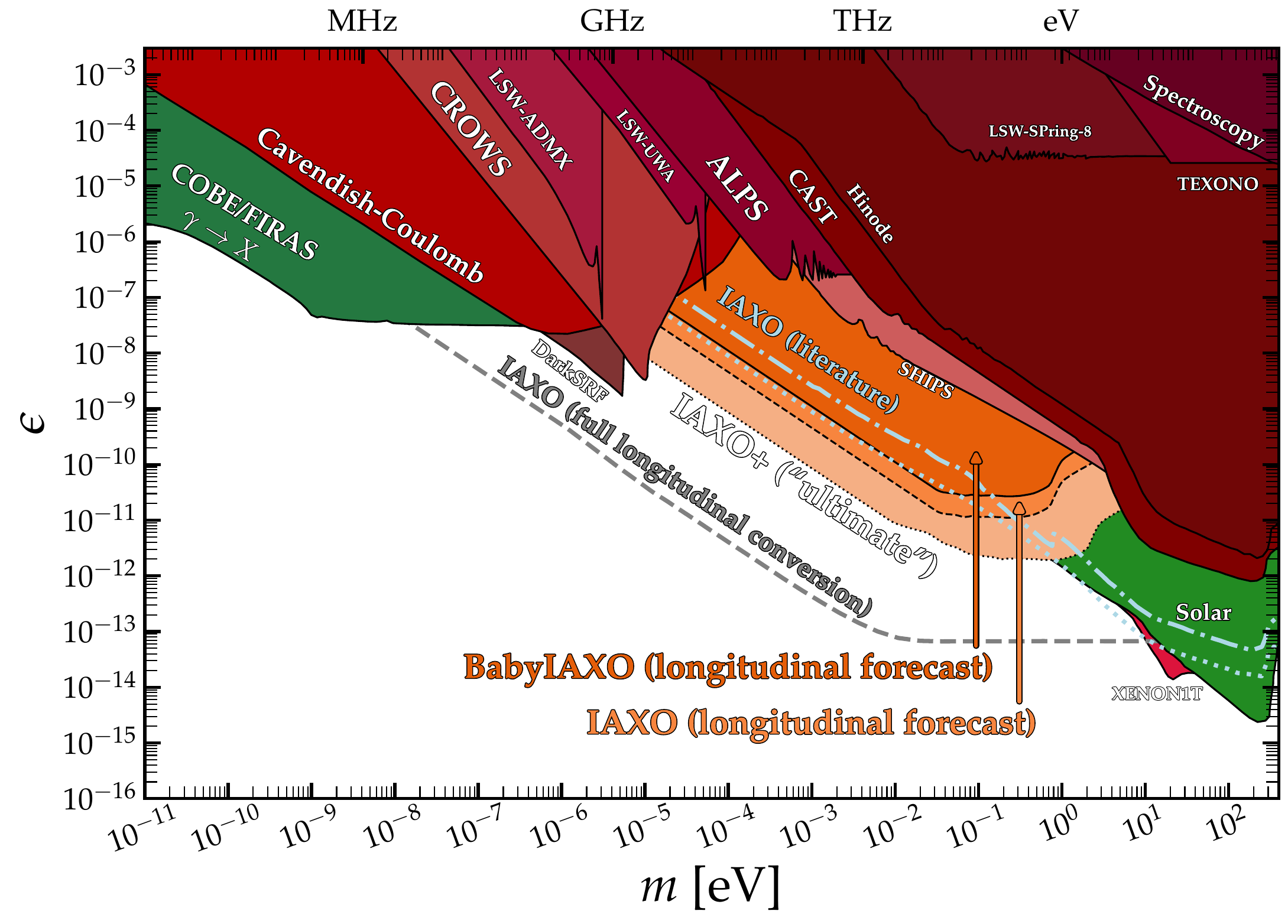}
    \caption{Plot of the expected constraints from CAST and IAXO (different shades of orange) compared to other results from the literature. We further show the expected reach of BabyIAXO (IAXO+) to the transverse flux as a light blue dash-dotted (dotted) line (both labelled as ``IAXO (literature)'') and the hypothetical maximum reach of IAXO for fully efficient conversion of the longitudinal component \cite{OShea:2023gqn} as a grey dashed line.  The constraints are taken from refs.~\cite{Caputo:2021eaa, AxionLimits}.}
    \label{fig:bounds_full}
\end{figure}

The phenomenology of the DP depends also on how its mass is generated.
If the DP mass is generated by the St\"{u}ckelberg mechanism \cite{Stueckelberg:1938hvi, Ruegg:2003ps, Kors:2004dx}, the parameter space is fully defined by the DP mass $m$ and the kinetic mixing $\epsilon$. If instead the mass is generated by the Higgs mechanism, an additional dark Higgs can be important for the phenomenology as well \cite{Ahlers:2008qc, Batell:2009yf, Schuster:2009au, Redi:2022zkt}; it should further be noted that the St\"{u}ckelberg mechanism can be interpreted as the decoupling limit of a heavy dark Higgs boson. For simplicity, we will thus focus on the St\"{u}ckelberg case.

There is a rich programme of experimental searches for the DP, especially since results from searches for axion-like particles (ALPs) can often be reinterpreted for the DP \cite{Jaeckel:2010ni}. In the following, we will restrict the discussion to the approximate mass range $ [10^{-6}\,\text{eV}, 10^2\,\text{keV}]$ as motivated by the sensitivity of the approach outlined in this work. Light-shining-through-a-wall (LSW) experiments like ALPS \cite{Ehret:2010mh} (and ALPS II \cite{ALPSII:2025eri}), CROWS \cite{Betz:2013dza}, or DarkSRF \cite{Kalia:2025afc} set stringent laboratory bounds that are independent of astrophysical or cosmological assumptions over a wide mass spectrum \cite{VanBibber:1987rq, Cameron:1993mr, Jaeckel:2007ch, Redondo:2010dp, Graham:2014sha}. Furthermore, laboratory tests of Coulomb's law present another strong probe \cite{Williams:1971ms, Bartlett:1988yy}, with peak sensitivity at masses similar to those of DarkSRF at $m\sim 10^{-6}\,$eV. For even lower masses, the strongest constraints arise from resonant photon-DP transitions at various stages of cosmological evolution, most notably from spectral distortions of the cosmic microwave background \cite{Mirizzi:2009iz, Kunze:2015noa, Caputo:2020bdy, Arsenadze:2024ywr, Chluba:2024wui}. For heavier DPs, i.e.\ between $m\sim 4\cdot 10^{-5}\,$eV and $m\lesssim 100\,$keV, astrophysical environments become particularly relevant: the non-observation of anomalous cooling in e.g.\ the Sun \cite{Redondo:2008aa, Redondo:2013lna} or globular cluster stars \cite{Dolan:2023cjs} gives rise to extremely strong bounds for masses larger than $\sim10\,$eV. Furthermore, traditional dark matter direct detection experiments are sensitive to the solar DP flux via absorption \cite{An:2013yua, An:2020bxd, XENON:2021qze}, providing the currently strongest bounds between $m\sim 4\cdot10^{-5}\,$eV and $m\sim 30\,$eV from XENON1T.\footnote{Technically, the solar cooling bounds are currently marginally stronger for $1\,\text{eV}\lesssim m \lesssim 5\,\text{eV}$.} 

The DP is also a well-motivated dark matter (DM) candidate, with a variety of production mechanisms \cite{Redondo:2008ec, Nelson:2011sf, Arias:2012az, Graham:2015rva}. The assumption that the DP is DM opens the door to many additional constraints from haloscopes \cite{Horns:2012jf,  Chaudhuri:2014dla}, direct detection \cite{An:2014twa, Bloch:2016sjj, Hochberg:2016sqx,XENON:2026gxb}, astrophysics \cite{Dubovsky:2015cca, Wadekar:2019mpc}, and cosmology \cite{Arias:2012az, McDermott:2019lch, Witte:2020rvb, Garcia:2020qrp}.\footnote{Note however a recent work \cite{Hook:2025pbn} which claims that many of the cosmological bounds are significantly weaker than expected.} We refer to ref.~\cite{AxionLimits} for a comprehensive and up-to-date list of DP probes, including DM searches. However, our results will apply independently of the DP being DM.

So far, we have ignored the so-called helioscope searches for DPs, a concept originally proposed for axions \cite{Sikivie:1983ip}. These instruments search for the solar flux of DPs: a Sun-tracking evacuated tube gives the (transverse component of the) DP time to oscillate into SM photons, which are then detected at the end of the tube \cite{Redondo:2008aa}. As the DP is massive, it has a physical longitudinal component, unlike the massless photon with only two transverse polarisations. Plasmons in the solar plasma allow for the production of longitudinal DPs, but inside the helioscope, which typically operates in a vacuum,\footnote{See however e.g.\ ref.`\cite{CAST:2013bqn} for an example of a helioscope filled with a buffer gas.} there is no SM analogue for the DP to oscillate into, rendering helioscopes essentially blind to the longitudinal flux. Thus, ever since it was discovered in 2013 that the longitudinal flux of DPs is much higher than the transverse part at masses $m\lesssim 1\,$eV \cite{An:2013yfc, Redondo:2013lna}, the limits have been dominated by either solar cooling or absorption in direct detection experiments \cite{An:2013yua}. 

In this work, we show that this limitation can be overcome: longitudinal DPs can be converted into photons at the surface of the mirrors in the X-ray optics of helioscopes. The conversion is a consequence of the electromagnetic (EM) interface conditions in the presence of a massive DP; these cannot be satisfied by a longitudinal wave alone and thus induce an additional outgoing transverse photon wave, in analogy to DP conversion at conducting surfaces in dish-antenna experiments \cite{Horns:2012jf, Jaeckel:2013sqa}. We derive the resulting conversion probability and apply it to the solar flux of longitudinal DPs for CAST \cite{CAST:2002aym}, SHIPS \cite{Schwarz:2015lqa} and the upcoming IAXO \cite{IAXO:2012eqj},
finding that IAXO will be able to probe previously unexplored parameter space beyond the direct detection constraints, even in the earlier BabyIAXO configuration. For convenience, we already show the results in figure~\ref{fig:bounds_full}.

This work is structured as follows: in section~\ref{sec:reflection}, we derive the conversion probability on the reflective surface of a perfect conductor under grazing incidence in detail and briefly touch upon the production of the solar DP flux. We then apply this knowledge to several different helioscope configurations in section~\ref{sec:helioscopes}. These results are then discussed in section~\ref{sec:disc} before wrapping up with the conclusions in section~\ref{sec:conc}. For the interested reader, we supply appendices with technical details. Throughout this paper, we work in natural units $\hbar=c=1$.

\section{Conversion of longitudinal dark photons on reflecting surfaces}\label{sec:reflection}

Let us begin by establishing the usual DP Lagrangian
\begin{align}
    \mathcal{L}&= -\frac{1}{4}F_{\mu\nu}F^{\mu\nu}-\frac{\epsilon}{2}F^\prime_{\mu\nu}F^{\mu\nu}-\frac{1}{4}F^\prime_{\mu\nu}F^{\prime\mu\nu}
    +\frac{m^2}{2}A^\prime_\mu A^{\prime\mu}-J_\text{EM}^\mu A_\mu\;,
\end{align}
where the first three parts are the kinetic terms, including the kinetic mixing, and the remaining parts cover the DP mass term and coupling of the photon to the EM current. We have expressed the (dark) vector potential and the (dark) field strength tensor as $A_\mu^{(\prime)}$ and $F_{\mu\nu}^{(\prime)}$, respectively. Here, we are using the mostly negative convention for the metric $g_{\mu\nu}=\text{diag}(1,-1,-1,-1)$.

Due to its non-vanishing mass, the DP, unlike its massless SM counterpart, carries a longitudinal component. Since most of this discussion takes place in vacuum, it is useful to perform a field redefinition that brings us into the so-called mass eigenbasis
\begin{align}\label{eq:rot}
    A_\mu&\to A_\mu-\frac{\epsilon}{\sqrt{1-\epsilon^2}} A_\mu^\prime\\
    A^\prime_\mu &\to \frac{1}{\sqrt{1-\epsilon^2}}A^\prime_\mu\\
    \Rightarrow \mathcal{L} &\to -\frac{1}{4}F_{\mu\nu}F^{\mu\nu}-\frac{1}{4}F^\prime_{\mu\nu}F^{\prime\mu\nu}+\frac{m^2}{2(1-\epsilon^2)}A^\prime_\mu A^{\prime\mu}-J_\text{EM}^\mu\left( A_\mu-\frac{\epsilon}{\sqrt{1-\epsilon^2}}A^\prime_\mu \right)\;.
\end{align}
From now on, we will only keep the lowest order in $\epsilon\ll1$.
For the classical equations of motion, we find
\begin{align}
    \left(\Box + \begin{pmatrix} m^2 & 0 \\ 0 & 0\end{pmatrix}\right)\begin{pmatrix}
        A_\mu^\prime \\ A_\mu
    \end{pmatrix}-\partial_\mu\partial^\nu\begin{pmatrix}
        A_\nu^\prime \\ A_\nu
    \end{pmatrix}=\begin{pmatrix}
       -\epsilon \\ 1
    \end{pmatrix} J_{\text{EM},\mu}\;.
\end{align}
Taking another derivative $\partial_\mu$ shows us, together with current conservation, that $\partial^\mu A^\prime_\mu=0$, i.e.\ the existence of the DP mass enforces the Lorenz condition which is \emph{not} a gauge choice. On the other hand, for the photon, we work in Lorenz gauge to arrive at an analogous expression. We can thus conclude that a DP wave (arriving e.g.\ from the Sun) can be expressed as 
\begin{align}
    A^\prime_\mu=X_{\mu}e^{-iP x}=X_{\mu}e^{-i(\omega t - \mathbf{p}\cdot\mathbf{x})}\;.
\end{align}
Enforcing the (pseudo-)Lorenz condition yields the relation $\omega X^0=\mathbf{p}\cdot\mathbf{X}$. Note that the right-hand side vanishes for transversely polarised DPs, which implies that only the spatial vector contains information on the wave. This is not true for the longitudinal case, where (except for non-relativistic DPs) the spatial and temporal components are similar in absolute value.

In the following, we will focus solely on the latter case and ignore the transverse components. We thus fix $X^\mu=X e^\mu_L$ with $e^\mu_L=1/m\,(p, \omega \hat{\mathbf{e}}_i)$, where $\hat{\mathbf{e}}_i$ denotes the unit vector in the direction of the incoming wave and $p=\sqrt{\omega^2-m^2}$. A simple calculation of the electric and magnetic fields yields
\begin{align}
    \mathbf{E}^\prime&=-\nabla (A^0)^\prime -\partial_t \mathbf{A}^\prime=\frac{\left(-ip^2 X +i \omega^2 X\right)}{m}\hat{\mathbf{e}}_i e^{-i(\omega t - \mathbf{p}\cdot\mathbf{x})} = i m X\hat{\mathbf{e}}_i e^{-i(\omega t - \mathbf{p}\cdot\mathbf{x})}\\
    \mathbf{B}^\prime &= \nabla\times \mathbf{A}^\prime=0\;.\label{eq:b_long}
\end{align}
Naturally, the magnetic component vanishes for the longitudinal mode.\footnote{Note that this seems to imply that the DP wave transports no energy as the Poynting vector vanishes, $\mathbf{E}\times\mathbf{B}=0$. However, due to the additional mass term, Poynting's theorem has to be modified; see appendix~\ref{app:poynting}.}

We further note that this mass eigenstate has a small interacting component. To account for this, we introduce a new vector notation as in ref.~\cite{Horns:2012jf}. The upper component denotes the sterile component, and the lower component is the interacting component. In this notation, we can write
\begin{align}\label{eq:Einc}
    \mathbf{E}^\prime=\begin{bmatrix}
        \mathbf{E}^\prime_\text{sterile}\\\mathbf{E}^\prime_{\text{int}}
    \end{bmatrix}=\begin{bmatrix}
        1\\-\epsilon
    \end{bmatrix}imX\hat{\mathbf{e}}_i e^{-i(\omega t - \mathbf{p}\cdot\mathbf{x})}\;.
\end{align}
This notation accounts for the fact that the wave couples with full strength to a hypothetical particle with dark charge (which we do not cover here) and with $\epsilon$-suppressed strength to EM charges and currents.

\begin{figure}[!t]
\centering
\def\sc{1.25}                              
\def\seg{10}                               
\begin{tikzpicture}[scale=\sc,
    wave/.style={-Stealth, decorate,
        decoration={snake, amplitude=1.2pt, segment length=\seg pt, post length=3pt}}]

\def\H{2.75}                              
\def\a{30}                                
\def\g{10}                                
\pgfmathsetmacro{\tg}{2*\a}               
\pgfmathsetmacro{\tr}{2*\a+\g}            
\pgfmathsetmacro{\tn}{\a+90}              
\pgfmathsetmacro{\Lg}{\H/sin(\tg)}        
\pgfmathsetmacro{\Lr}{\H/sin(\tr)}

\pgfmathsetmacro{\hw}{\seg/2/28.4528/\sc}
\pgfmathsetmacro{\rI}{6*\hw}              
\pgfmathsetmacro{\rA}{5*\hw}              
\pgfmathsetmacro{\rO}{10*\hw}             

\coordinate (M) at (6,0);                 
\coordinate (A) at ($(M)+(\a+180:2)$);    
\coordinate (B) at ($(M)+(\a:2)$);

\pgfmathsetmacro{\Ldp}{\seg/2+\seg*ceil((5*28.4528*\sc-\seg/2)/\seg)}   
\pgfmathsetmacro{\xe}{6-\Ldp/(28.4528*\sc)}             
\draw[wave, Stealth-, decoration={pre length=\seg/2 pt, post length=0pt}] (M) -- (\xe,0);
\node[above=3pt] at (3.5,0) {DP};

\shade[inner color=yellow!50, outer color=yellow!60!orange] (0,0) circle (1);
\node[below=2pt] at (0,-1) {Sun};

\draw[blue, thick] (A) -- (B);
\foreach \i in {1,...,19} {
    \draw[gray, thin] ($(A)!{\i/20}!(B)$) -- ++(\a-45:0.25);
}

\draw[-Stealth] (M) -- ++(\tn:1.5);

\draw[wave, gray] (M) -- ++(\tg:\Lg);
\draw[wave, red]  (M) -- ++(\tr:\Lr) node[pos=0.8, left=3pt, text=black] {$\gamma$};

\draw ($(M)+(180:\rI)$) arc (180:{180+\a}:\rI);
\node at ($(M)+({180+\a/2}:{\rI+0.4})$) {$\alpha$};

\pgfmathsetmacro{\Lt}{\seg*ceil(3*28.4528*\sc/\seg)/(28.4528*\sc)}   
\draw[wave, black, decoration={pre length=\seg/2 pt, post length=\seg/2 pt}] (M) -- ++(\Lt,0);

\draw ($(M)+(\a:\rA)$) arc (\a:\tg:\rA);
\node at ($(M)+({(\a+\tg)/2}:{\rA+0.3})$) {$\alpha$};

\draw ($(M)+(\a:\rO)$) arc (\a:\tr:\rO);
\node at ($(M)+({\a+12}:{\rO+0.3})$) {$\beta$};

\node[fill=black, text=white, minimum width=2cm, minimum height=1cm,
      anchor=south] at (7.3,\H) {Detector};

\end{tikzpicture}
\caption{Schematic representation of the interconversion process of solar DPs in the telescope: the Sun emits a flux of DPs (black line), which can be converted on a reflecting surface (blue) with relative angle $\alpha$. The outgoing photon flux (red) is emitted at a larger angle $\beta\geq\alpha$. The grey line indicates the direction of ordinary specular reflection for comparison. Note that most of the energy of the incoming DP wave passes through the mirror, as indicated by the continued black line. In a realistic setup, the angles are much smaller.}
\label{fig:signals}
\end{figure}
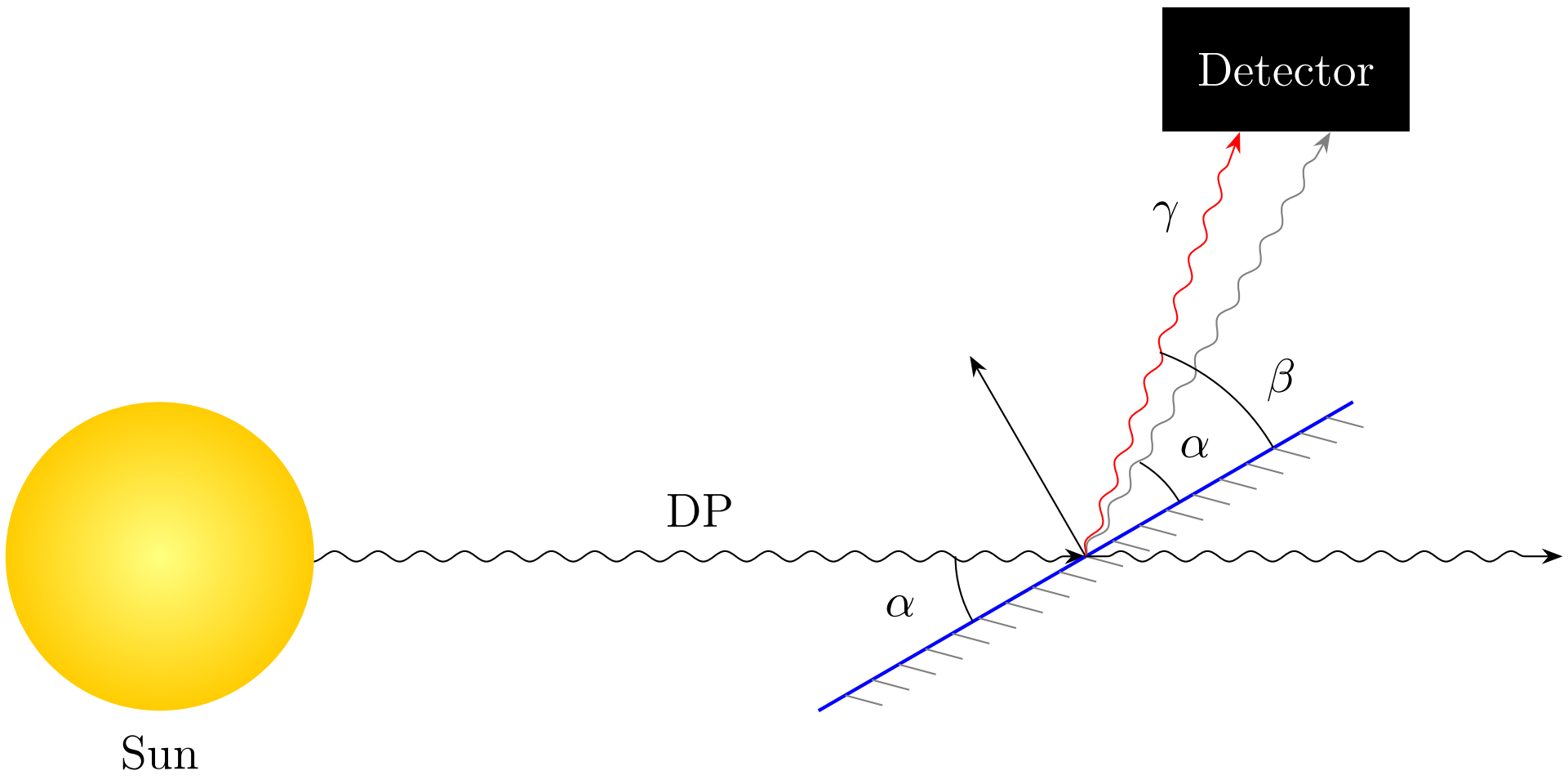

From electromagnetism, we know that the interacting component parallel to an ideally conducting surface has to vanish.\footnote{To be precise, the parallel component cannot have a discontinuity, and thus if the field vanishes inside the conductor, it has to vanish on the outside as well.} Let us therefore imagine that a wave of DPs impinges on such a conductor under a fixed angle $\alpha$ (we measure the angle between the surface of the conductor and the direction of propagation, i.e.\ a small angle is almost parallel propagation). In this scenario, the field on the surface has to vanish. Let $\mathbf{x}_s$ denote the points on the surface and $\hat{\mathbf{n}}$ the normal vector to the surface. We then find
\begin{align}
    \hat{\mathbf{n}}\times \mathbf{E}_\text{tot}=\hat{\mathbf{n}}\times\begin{bmatrix}
        \mathbf{E}^\prime_\text{sterile}\\0
    \end{bmatrix}_{\mathbf{x}=\mathbf{x}_s }\;.
\end{align}
In this suggestive way of writing, we have already made clear that the DP wave alone cannot satisfy this condition. We need an additional \emph{outgoing} wave to compensate for the DP wave passing through. In particular, this wave must be mostly electromagnetic, as the sterile component of the DP wave does not feel the presence of the wall at leading order. For better visualisation, we present a sketch of the setup in question in figure~\ref{fig:signals}. We thus make the ansatz
\begin{align}
    \mathbf{E}_\text{tot}=\mathbf{E}^\prime+\mathbf{E}_\text{out}=\begin{bmatrix}
        1\\-\epsilon
    \end{bmatrix}imX \hat{\mathbf{e}}_i e^{-i(\omega t - \mathbf{p}\cdot\mathbf{x})}+\begin{bmatrix}
        \epsilon\\1
    \end{bmatrix}m\mathbf{X}_\text{out}e^{-i(\omega t - \mathbf{k}\cdot\mathbf{x})}\;.
\end{align}
This ansatz already contains energy conservation, i.e.\ the outgoing wave must have the same frequency as the incoming one.\footnote{Due to the breaking of spatial homogeneity, momentum is not conserved at the boundary.} We have already accounted for the fact that the outgoing particle will be the mostly interacting, massless eigenstate and thus $\mathbf{p}\neq\mathbf{k}$ but rather $\mathbf{k}^2=\omega^2$. Now, we have to demand that the interacting component vanishes on the surface. This provides us with the equation
\begin{align}
    \hat{\mathbf{n}}\times\left( i\epsilon m X \hat{\mathbf{e}}_i e^{-i(\omega t - \mathbf{p}\cdot\mathbf{x}_s)}-m\mathbf{X}_\text{out}e^{-i(\omega t - \mathbf{k}\cdot\mathbf{x}_s)}\right)\overset{!}{=}0\;.
\end{align}
As this has to be valid on the entire surface, we can solve for the phase and the amplitude separately. Let us begin with the latter
\begin{align}
    X_{\text{out},\parallel}=i\epsilon X\cos\alpha\;.
\end{align}
This condition fixes the component parallel to the surface completely. The other condition for the field to vanish on the surface is
\begin{align}
    k_\parallel=p\cos\alpha\;.
\end{align}
As the photon is massless, we can find the perpendicular momentum as
\begin{align}
    \mathbf{k}^2&=k_\parallel^2+k_\perp^2=\omega^2\\
    \Rightarrow k_\perp&=-\sqrt{\omega^2-(\omega^2-m^2)\cos^2\alpha}=-\sqrt{\omega^2\sin^2\alpha+m^2\cos^2\alpha}\;,
\end{align}
where the negative sign accounts for the actual reflection of the wave.
In particular, this allows us to calculate the outgoing angle
\begin{align}
    \cos\beta=\frac{k_\parallel}{|\mathbf{k}|}=\sqrt{1-\frac{m^2}{\omega^2}}\cos\alpha\;,\label{eq:angle}
\end{align}
which indicates that for relativistic DPs, as most of the solar flux would be, the outgoing angle is extremely close to the one expected from the normal EM reflection with $\beta\simeq\alpha$.\footnote{On the other hand, for non-relativistic DPs the reflection is essentially always perpendicular to the mirror. This fact was observed (independently of the polarisation) and used in ref.~\cite{Horns:2012jf} for DP dark matter searches.}

Our last exercise is now to calculate the perpendicular component of the amplitude. For this, we can make use of Maxwell's equations (in vacuum)
\begin{align}
    \nabla\cdot \mathbf{E}_\text{out}&\overset{!}{=}0\Rightarrow k_\perp X_{\text{out},\perp}=-k_\parallel X_{\text{out},\parallel}\\
    \Rightarrow X_{\text{out},\perp}
    &=i\frac{\sqrt{\omega^2-m^2}}{\sqrt{\omega^2\sin^2\alpha+m^2\cos^2\alpha}}\epsilon X\cos^2\alpha\;.
\end{align}
To determine the energy flow, we will calculate the (time-averaged) Poynting vector (for a massless vector). Therefore, our next step is to infer the magnetic field component. For simplicity, we will focus only on the interacting part
\begin{align}
    \partial_t \mathbf{B}_\text{out}&=-\nabla \times \mathbf{E}_\text{out}\\
    \Rightarrow \mathbf{B}_\text{out}&=\frac{m}{\omega}(\mathbf{k}_\parallel+\mathbf{k}_\perp)\times\left(\mathbf{X}_{\text{out},\parallel}+\mathbf{X}_{\text{out},\perp}\right)e^{-i(\omega t - \mathbf{k}\cdot\mathbf{x})}\\
    &= -i\epsilon\frac{m}{\omega}X\cos\alpha\Bigg(\frac{(\omega^2-m^2)\cos^2\alpha}{\sqrt{\omega^2\sin^2\alpha+m^2\cos^2\alpha}}+\sqrt{\omega^2\sin^2\alpha+m^2\cos^2\alpha}\Bigg)e^{-i(\omega t - \mathbf{k}\cdot\mathbf{x})}\hat{\mathbf{e}}_B
\\
    &=-i\epsilon\frac{\omega m}{\sqrt{\omega^2\sin^2\alpha+m^2\cos^2\alpha}}Xe^{-i(\omega t - \mathbf{k}\cdot\mathbf{x})}\cos\alpha\hat{\mathbf{e}}_B
    \;.
\end{align}
Here we have introduced a new unit vector $\hat{\mathbf{e}}_B$ which indicates the direction of the B-field, orthogonal to the electric component and the direction of propagation of the outgoing wave. Since we already know the direction of propagation $\hat{\mathbf{e}}_k$, we can just estimate the time-averaged flux of EM energy via
\begin{align}
\langle\mathbf{S}\rangle&=\frac{\text{Re}\left(\mathbf{E}_\text{out}\times\mathbf{B}^\ast_\text{out}\right) }{2}=\frac{1}{2}\epsilon^2\frac{\omega^2 m^2}{\omega^2\sin^2\alpha+m^2\cos^2\alpha}\cos^2\alpha X^2 \hat{\mathbf{e}}_k\\
    \Rightarrow|\langle\mathbf{S}\rangle| & = \frac{1}{2}\epsilon^2\frac{\omega^2 m^2}{\omega^2\sin^2\alpha+m^2\cos^2\alpha}\cos^2\alpha X^2 \;.
\end{align}
Of course, the reflected flux is in the direction of $\mathbf{k}$, which can be almost perpendicular to the surface for non-relativistic DPs. In the helioscope, we have to account for the fact that photons too strongly deviating from the usual reflection angle will not reach the detector.

This flux has to be contrasted with the incoming flux of energy. For the massive DP, the flux of energy has a second component which depends directly on the potentials and not the fields (cf.\ appendix~\ref{app:poynting}). The expression of relevance is
\begin{align}
    \langle \mathbf{S}^\prime\rangle&=\frac{m^2}{2}\text{Re}\left(A_0^\prime \mathbf{A}^{\prime\ast}\right)\Rightarrow |\langle\mathbf{S}^\prime\rangle|=\frac{\omega^2}{2}\sqrt{1-\frac{m^2}{\omega^2}}X^2=\frac{\omega^2}{2}\frac{\cos\beta}{\cos\alpha}X^2\;.
\end{align}
In conclusion, we can define an effective interconversion probability as
\begin{align}\label{eq:prob}
    \mathcal{P}_0=\frac{ |\langle\mathbf{S}\rangle|}{|\langle\mathbf{S}^\prime\rangle|}
    &=\epsilon^2 \frac{m^2\cos^2\alpha}{\left(\omega^2\sin^2\alpha+m^2\cos^2\alpha\right)\sqrt{1-\frac{m^2}{\omega^2}}}\\
    &\simeq\epsilon^2\begin{cases}
        \frac{m^2}{\omega^2}\cot^2\alpha\;,&\quad \omega \alpha \gg m\\
        \frac{1}{\sqrt{1-\frac{m^2}{\omega^2}}}\;,&\quad \omega \alpha \ll m\\
    \end{cases}\;.
\end{align}
Note that this result seemingly diverges for non-relativistic DPs so that $\mathcal{P}_0>1$ for very slow DPs. However, this result is an unphysical relic of the expansion in small $\epsilon$ and the expression should only be used when $\mathcal{P}_0\ll 1$.\footnote{Note that without this enhancement, the probability is always smaller than $\epsilon^2\ll 1$ so that, in the parameter space of interest, it is always much smaller than 1, even for very non-relativistic DPs.} In the next section, we will see that the incoming DP flux is suppressed by a kinematic $\sqrt{1-m^2/\omega^2}$, i.e.\ its velocity, which renders the outgoing photon flux finite, even in the non-relativistic limit.

\subsection{The longitudinal solar DP flux}\label{sec:solarflux}

As mentioned in the introduction, in early works the longitudinal flux was underestimated. While this was of no concern for the conventional helioscope searches \cite{Redondo:2008aa}, it led to a miscalculation of the stellar cooling power of the DP. The so-called solar cooling argument relies on the DP luminosity being only a fraction of the total solar luminosity. This led to the conclusion that helioscope bounds are significantly stronger than cooling bounds. However, once this error was noted and improved calculations were available \cite{An:2013yfc,Redondo:2013lna}, the opposite was the case: the resulting solar energy loss constraints were substantially stronger in the regime of light DPs due to the enhanced emission of longitudinal DPs. Shortly after that, it was realised that liquid xenon detectors can directly absorb the longitudinal component \cite{An:2013yua}, leading to even stronger bounds \cite{An:2013yua, An:2020bxd, XENON:2021qze}.

A detailed calculation of the solar DP flux is beyond the scope of this work, as we want to focus on the new mode of detection. We will therefore consult ref.~\cite{Redondo:2013lna} to extract the longitudinal DP flux.\footnote{To be precise, we use figure~4 for the resonant regime and use their interpolation formula eq.~(4.11) for higher energies.} To obtain the flux for a non-zero DP mass, we multiply the flux by $\sqrt{1-m^2/\omega^2}$ as mentioned before.

\begin{figure}
    \centering
    \includegraphics[width=\linewidth]{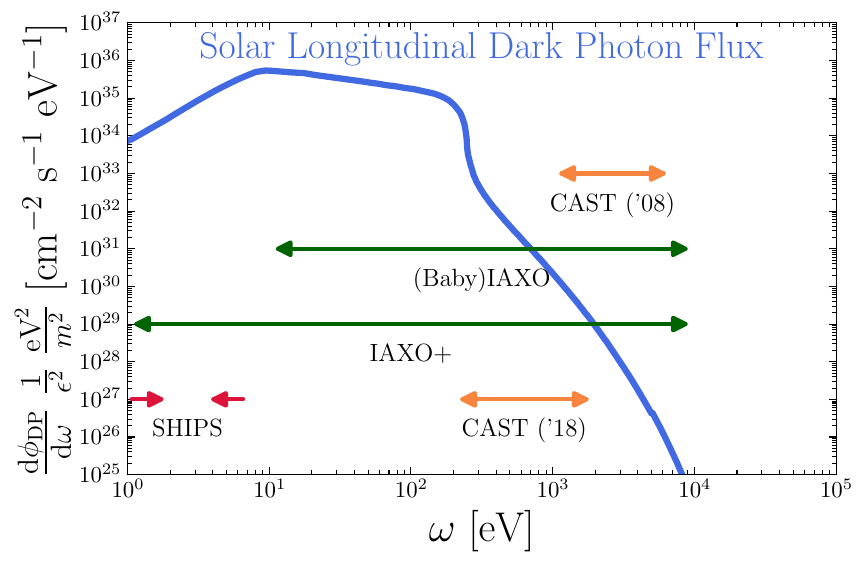}
    \caption{Plot of the solar longitudinal DP flux (blue). Note that for a given DP mass $m$, the flux needs to be multiplied by $\sqrt{1-m^2/\omega^2}$. We also show the (projected) detection range of different helioscopes: CAST analyses are shown in orange, SHIPS in red, and IAXO in green.}
    \label{fig:flux}
\end{figure}

To aid the reader, we show this result in figure~\ref{fig:flux} modulo the factor $\sqrt{1-m^2/\omega^2}$. The flux peaks at UV energies and, for light DPs, scales $\propto m^2$.

Ref.~\cite{Redondo:2013lna} provides the flux in units of cm$^{-2}$s$^{-1}$keV$^{-1}$. To determine the outgoing photon flux, we thus use 
\begin{align}\label{eq:eff_flux}
    \frac{\text{d}\phi}{\text{d}\omega}=\epsilon^2 \frac{m^2\cos^2\alpha}{\left(\omega^2\sin^2\alpha+m^2\cos^2\alpha\right)}\frac{\text{d}\phi_\text{DP}}{\text{d}\omega}=\epsilon^4\frac{\text{d}\tilde{\phi}}{\text{d}\omega}\;,
\end{align}
where we have already cancelled the divergent term with the extra $\sqrt{1-m^2/\omega^2}$ from the incoming flux to arrive at a finite outgoing flux.\footnote{Of course, this does not mean that the flux is finite for very non-relativistic DPs. The apparent finiteness comes from the so-called traffic jam effect where, due to their low velocity, a large number density of DPs can build up inside the telescope; the potential well of the Sun will prevent these slow DPs from ever reaching the detector in practice.} In the next section, we will discuss specific examples and comment on the relevance of the optics of a given instrument to the sensitivity to the converted photons; furthermore, we will discuss the effect of finite conductivity of the mirror for grazing incidence.
We note in passing that the flux has an energy-dependent angular distribution: for energies below $\sim 0.3\,$keV, the flux is dominated by resonant production in the layer where the local plasma frequency matches the energy, $\omega=\omega_p(r)$ \cite{An:2013yfc, Redondo:2013lna}. Since $\omega_p$ decreases outwards, low-energy DPs originate from increasingly outer layers, e.g.\ $r\simeq0.8\,R_\odot$ for $\omega\simeq10\,$eV, whereas the non-resonant keV flux stems predominantly from the hot core.

\section{Looking at different helioscopes}\label{sec:helioscopes}

In this section, we will briefly discuss how (X-ray) helioscopes can be used in combination with the previous result. The typical X-ray reflectivity, even for good conductors, is relatively poor. Therefore, the principle of grazing incidence is used. In the X-ray regime, the typical real part of the refractive index of metals is slightly below 1, $n=1-\delta+i\beta_n$. In that case, incoming light can experience total external reflection under sufficiently small angles \cite{B_Aschenbach_1985}. For small $\delta$, one can find the relation $\alpha_t=\sqrt{2\delta}$, where $\alpha_t$ is the angle below which total reflection occurs. Ref.~\cite{B_Aschenbach_1985} also provides a simple parametric formula (valid for heavy elements) which estimates this angle in rad as
\begin{align}\label{eq:angle_est}
    \alpha_t=0.064\left(\frac{1\,\text{keV}}{\omega}\right)\left(\frac{\rho}{10\,\text{g/cm}^3}\right)^{1/2}\;.
\end{align}
We see that this angle is naturally very small in the X-ray regime. We note in passing that in the cold plasma limit $\delta\simeq\omega_p^2/(2\omega^2)$ with $\omega_p$ the plasma frequency of the reflector. 

We should emphasise that the reflector is not a perfect conductor in this setup; we provide a derivation of the conversion probability for ``total external DP reflection'' in appendix~\ref{app:ter}. It turns out that the case of longitudinal DPs leads to lower conversion probabilities. This is largely driven by the fact that the in-medium effective coupling is hardly changed for the longitudinal component at X-ray energies. In practice, we use the exact result 
\begin{align}
    \mathcal{P}
    &=\mathcal{P}_0\frac{\left|1-n^2\right|^2}{\left|n^2+\frac{\tan\beta_m}{\tan\beta}\right|^2}\;,\label{eq:prob_ex}
\end{align}
where we use $\mathcal{P}_0$ from eq.~\eqref{eq:prob} and $n$ is the (complex) refractive index. $\beta$ is the same outgoing angle as before (cf.\ section~\ref{sec:reflection}) and $\beta_m$ is defined from the dispersion relation of the photon in medium and derived in appendix~\ref{app:ter}. The relevant final result is
\begin{align}
    \cos\beta_m&=\frac{\sqrt{1-\frac{m^2}{\omega^2}}}{\sqrt{n^2}}\cos\alpha\;,
\end{align}
indicating that $\beta_m$ can be complex for general $n$ or for real $n^2$ sufficiently below 1.
It should be noted that eq.~\eqref{eq:prob_ex} contains our previous result eq.~\eqref{eq:prob} in the approximation of a perfect conductor. For simplicity, we assume a gold-coated mirror surface for all helioscopes.\footnote{BabyIAXO will use multilayer coatings inspired by NuSTAR \cite{Yu:2026ykr}. A detailed derivation of the conversion properties of this type of reflector is postponed to future work.} We further note that the conversion does not require ``total external reflection'', i.e.\ angles below $\alpha_t$, but it tends to increase the interconversion probability. We can just multiply the outgoing flux of eq.~\eqref{eq:eff_flux} by the factor $\mathcal{P}/\mathcal{P}_0$ of eq.~\eqref{eq:prob_ex}.

\subsection{How to do a (simple) analysis}

In the following subsections, we will typically quote upper bounds for fluxes over the energy range of a given experiment. We can then constrain the kinetic mixing parameter by demanding
\begin{align}\label{eq:eps_analysis}
    \epsilon(m) \lesssim \left(\frac{\frac{\text{d}\phi_\text{exp}}{\text{d}\omega}}{\underset{\omega_\text{exp}}{\max}\left[\frac{\text{d}\tilde{\phi}}{\text{d}\omega}(\omega, m)\right]}\right)^{1/4}\;,
\end{align}
with $\text{d}\phi_\text{exp}/\text{d}\omega$ being the experimental upper limit on the flux, which is flat in energy for many of the setups considered here, and $\omega_\text{exp}$ the experimental energy range.\footnote{The energy with the largest contribution is usually determined by the energy dependence of the flux and the instrument's quantum efficiency, where the former is highest at the lowest available energies for X-ray energy DPs.}

\subsection{CAST}
Let us discuss the CAST experiment as a prime example of such an instrument. For this, we will take inspiration from the DP analysis of ref.~\cite{Redondo:2008aa}; however, we now focus on the longitudinal component and redo some of the steps.

CAST was a helioscope designed to search for solar axions.\footnote{More recently, it has also been used as a haloscope in a different setup \cite{Melcon:2018dba,CAST:2020rlf}.} To this end, a Wolter-I X-ray telescope was attached to a decommissioned LHC prototype dipole magnet tube. The idea was to convert solar axions in the keV range into photons in the magnetic field of the long tube. These would then be collected in the telescope attached to the end of the instrument. The focusing of the converted photons was found to significantly improve the signal-to-noise ratio \cite{Kuster:2007ue}. We note here that a Wolter-I telescope employs first a reflection on a paraboloidal mirror followed by a second reflection on a hyperboloidal mirror to efficiently focus X-rays. In our setup, the first mirror is responsible for the interconversion, while the second mirror works as intended for conventional telescopes.

Before we start inferring the bounds, we should, however, apply some cuts on the energy range. In particular, if the particle is not sufficiently relativistic, the mirror will not properly reflect the converted photons onto the detector. We will use two different assumptions: the conservative choice is to demand that the difference in angle lies below the angular resolution quoted in ref.~\cite{Kuster:2007ue}: $\alpha_\delta=34.5\,$arcsec$\,\simeq 1.7\cdot 10^{-4}$. This resolution was found for $\omega=1.5\,$keV, but we assume that it is a good approximation of the resolution for other energies as well. However, we do not necessarily have to care about the resolution of the ``DP image of the Sun'' as long as all photons make it through the telescope into the detector. A more optimistic choice would thus be to just demand that the relative deviation of the angle is at most $\mathcal{O}(1)$. To make sure that a photon from a DP conversion hits the detector where a photon from conventional reflection would hit the detector, we should consult eq.~\eqref{eq:angle}. It is already quite clear that we can assume an ultra-relativistic setup, as otherwise the discrepancy would automatically be enormous (except for perfectly perpendicular reflection which is of no interest to us here). Writing $\beta=\alpha +\delta \alpha$ and then expanding in small $m/\omega$, we find
\begin{align}
    \delta\alpha\simeq \frac{1}{2}\frac{m^2}{\omega^2}\cot\alpha\xrightarrow{\alpha\to 0}\frac{1}{2}\frac{m^2}{\omega^2\alpha}\;,
\end{align}
i.e.\ the outgoing angle is always larger than the incoming one.
Once more, we recover the condition $m<\omega\alpha$ so that the deviation from the conventional reflection angle also depends on the incoming angle. For our two choices ($\delta \alpha \lesssim \alpha_\delta $ and $\delta\alpha\lesssim\alpha$), we find the cuts at
\begin{align}\label{eq:cuts_CAST}
    \frac{m}{\omega}\overset{!}{\lesssim}  \begin{cases} \,  \sqrt{\alpha\alpha_\delta}\simeq 0.001\, ,\quad &\quad \text{conservative}\\
   \quad\quad \alpha \simeq 0.009\, ,\quad &\quad \text{optimistic}\end{cases}\;,
\end{align}
where we conservatively neglected a factor $\sqrt{2}$.
To estimate the incoming angle, we use the relation between focal length $L_\text{f}=1600\,$mm and the shell radii $R_\text{a}=38\,$--$\,81.5\,$mm (depending on the shell) \cite{CAST:2007jps} which reads $\alpha\simeq R_\text{a}/(4L_\text{f})\sim 0.006\,$--$\,0.013\,$ \cite{Armengaud:2014gea}. For simplicity, we just present the central value $\alpha\simeq 0.009$ in this case.
We see that truly only very relativistic DPs fulfil this criterion. We will briefly explore a tuned setup which could avoid this issue later in section~\ref{sec:narrowband}.
We have backed up our findings with a toy simulation in appendix~\ref{app:raytracing}. For an IAXO-like setup (cf.\ the upcoming section~\ref{ssec:iaxo}), we found that the optimistic scenario is achievable.\footnote{The new IAXO results are the key findings of this paper and thus we focused on this setup for our simulation.} We assume that this finding carries over to a CAST-like configuration. 

Furthermore, we should account for the fact that the telescope is not perfectly reflecting and the detector not fully efficient. We should therefore weight the incoming flux by the ratio of effective area to the geometric area. The latter is quoted as $14.5\,$cm$^2$. The former depends on the energy and can be extracted from ref.~\cite{Kuster:2007ue}. The quantum efficiency of the detector is very close to one in the entire energy range explored by CAST. Ref.~\cite{Redondo:2008aa} bases its analysis on a background in the window [0.5\,keV,15\,keV] which was measured to be $\text{d}\phi_\text{CAST '08}/\text{d}\omega\lesssim (1$--$\,4)\cdot10^{-6}\,$keV$^{-1}$s$^{-1}$cm$^{-2}$, further introducing a fudge factor to account for possible systematics. It is unclear where exactly this number comes from; consulting the original CAST analysis \cite{CAST:2007jps}, we see that their figure~7 provides similar numbers. These are, however, measured by a TPC detector and \emph{not} by the telescope. We will thus extract the numbers from figure~6 of ref.~\cite{CAST:2007jps}: over an energy range from 1\,keV to 7\,keV with 20 equally spaced bins, and for a conservative (optimistic) bound, there should not be more than 4 (2) counts per bin for an exposure time of 197\,h. Together with a spot size of $9.35\,$mm$^2$ compared to the geometric area of $14.5\,$cm$^2$, this leads to an upper limit on the flux of $1\cdot 10^{-7}\,$keV$^{-1}$s$^{-1}$cm$^{-2}$ ($5\cdot 10^{-8}\,$keV$^{-1}$s$^{-1}$cm$^{-2}$). Furthermore, we plug $\alpha=0.013$ ($\alpha=0.006$) into eq.~\eqref{eq:eff_flux}, while we enforce the kinematic cuts according to the central value used for eq.~\eqref{eq:cuts_CAST}.

\begin{figure}
    \centering
    \includegraphics[width=\linewidth]{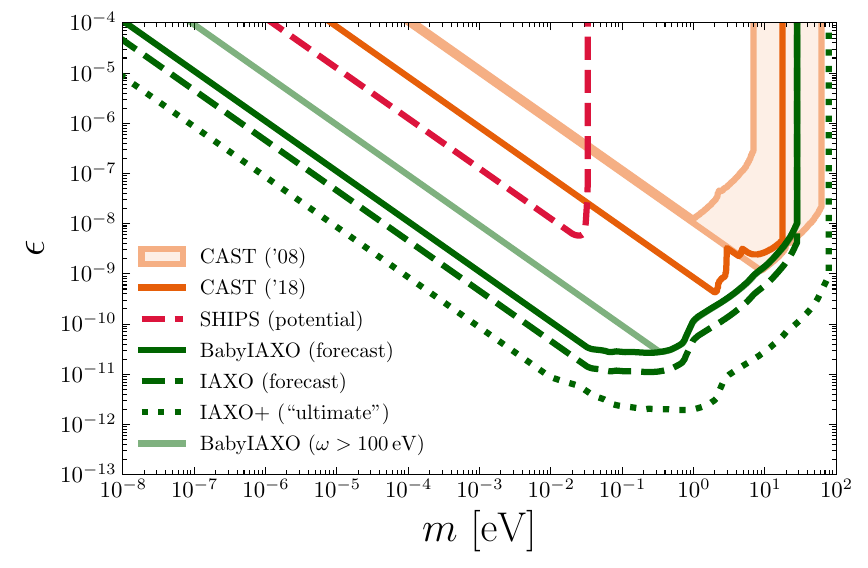}
    \caption{Plot of the expected constraints on the kinetic mixing $\epsilon$ as a function of the DP mass $m$. We show the results of our analysis for CAST in different shades of orange, with the band indicating the dependence of the reach on our assumptions for the '08 analysis. The potential reach of a SHIPS-like helioscope is shown as a red dashed line, and we show the sensitivity forecast of the BabyIAXO/IAXO/IAXO+ experiments as a solid/dashed/dotted dark green line. We further show the impact of the energy range for BabyIAXO as a solid light green line.}
    \label{fig:bounds}
\end{figure}

The result can be found in figure~\ref{fig:bounds}. We show the possible reach as an orange band with the upper boundary set by the conservative reach and the lower boundary by the optimistic case. Note that the difference between the conservative and more optimistic scenarios decreases significantly for lower masses ($m\lesssim 0.1\,$eV). Thus, in the regime where the longitudinal mode is strongest anyway, the bounds vary by a factor of $\sim 1.2$ and thus can be considered relatively robust against our assumptions.

Towards lower masses, the bounds decouple as $\epsilon  \propto 1/m $.
We should investigate this regime further: the solar longitudinal DP flux scales generically with $\epsilon^2 m^2$ as shown in figure~\ref{fig:flux} (see also ref.~\cite{Redondo:2013lna}), and the interconversion probability in eq.~\eqref{eq:prob_ex} also approximately scales like  $\epsilon^2m^2$ as long as $m\cos\alpha\ll \omega \sin\alpha$. Overall, the total flux then scales like $\epsilon^4m^4$, which explains the scaling.
The breakdown of this regime can be understood from the kinematic cuts: the bound arises from the lowest available energy $\omega\sim 1\,$keV where the DP flux is maximal in the given range (see figure~\ref{fig:flux}), the effective area is not far from its maximum value, and the conversion probability is also enhanced (cf.\ eq.~\eqref{eq:prob_ex}). This only changes when the condition of eq.~\eqref{eq:cuts_CAST} is violated; for the conservative (optimistic) bound, this happens at $m\sim 1\,\text{keV}\times 0.001=1\,$eV ($m\sim 1\,\text{keV}\times 0.009=9\,$eV), in excellent agreement with the break in the inversely proportional behaviour. From this point on, the kinematic cut pushes the energy window towards higher energies where the flux decreases. Eventually, it reaches a point where the required energies are outside the region where the effective area has support, and thus the bounds vanish where $m\sim 7\,\text{keV}\times 0.001=7\,$eV ($m\sim 7\,\text{keV}\times 0.009=63\,$eV). It should be noted that the restricted energy range is the biggest issue in this setup; if the sensitivity could be held towards lower energies, the bounds would improve strongly due to the increase in flux (cf.\ figure~\ref{fig:flux}).

\subsubsection{CAST operating in  ``chameleon mode''}

In 2013, CAST extended its programme to lower energies \cite{CAST:2015npk}, looking for so-called solar chameleons \cite{Brax:2010xq}. In particular, they targeted lower energies of $\mathcal{O}(100\,\text{eV})$ in this run. However, they did \emph{not} use a telescope in this setup, making it effectively insensitive to longitudinal DPs.

Later on, a more refined search result was presented in ref.~\cite{CAST:2018bce}, this time employing a telescope. We use the quoted observation time of $254\,$h and the quantum efficiency in their figure~2. We then use their figure~5 (assuming that their ``gold region'' is also where most of the longitudinal DP flux will be found; for a more detailed comment on the longitudinal DP signal region see section~\ref{ssec:iaxo}) to estimate an upper bound on the incoming flux. We observe that the signal should not exceed $\sim 5$ events per 200\,eV bin. Together with the observation time of 254\,h, a spot size of 25\,mm$^2$, and a geometric area of 14.5\,cm$^2$, we see that the photon flux due to converted DPs should not exceed $5\cdot 10^{-7}\,$keV$^{-1}$s$^{-1}$cm$^{-2}$ for the entire range $[200,2000]\,$eV.\footnote{Curiously, the '08 run of CAST was observing fewer events in a similar observation time.} To simplify the discussion, we just use the central value $\alpha=0.009$ for this analysis, and we stay in the optimistic scenario.

The result is presented again in figure~\ref{fig:bounds} as a dark orange line. We find the same behaviour in the low mass tail as before. Due to its superior energy range, the 2018 result clearly outperforms the other CAST bounds below $m = \mathcal{O}(\text{few})\,$eV. Only at higher masses, when the kinematic cuts become more important, does the sensitivity drop rather steeply as the quantum efficiency becomes small in the kinematically allowed region.

\subsection{Visible light regime and SHIPS' potential reach}

Let us also discuss the smaller SHIPS helioscope, which was dedicated to searching \emph{only} for DPs and thus did not contain the usual magnet \cite{Schwarz:2015lqa}. The instrument focused on the visible and near-UV range where --- before the realisation of the strong longitudinal cooling bound --- it was expected to set leading bounds for lighter DPs. Looking at the DP flux, it seems advantageous at first to operate in this regime where the flux is higher than in X-rays. However, the problem is that for these wavelengths, grazing incidence telescopes are typically not used.\footnote{The construction of SHIPS was ``piggybacking'' the \href{https://www.fundus.uni-hamburg.de/de/collection_records/93742}{Oskar L\"{u}hning Telescope} (website in German) and employed a simple lensing system to focus the converted photons onto a detector.}

Nevertheless, in principle it should be possible to also construct such a telescope for lower energy photons like visible light, especially because one can reach high reflectivity without relying on grazing incidence. We will thus discuss the potential of a helioscope with the specifications of SHIPS but with the optics inspired by CAST, assuming as before $\alpha=0.009$. We can convolve the solar DP flux with the conversion probability and effective area of the telescope using eq.~(4.5) of ref.~\cite{Schwarz:2015lqa}. This photon flux needs to be below $\text{d}N/\text{d}t=1.2\,$mHz. The signal rate is
\begin{align}
    \frac{\text{d}N_\text{SHIPS}}{\text{d}t}=A\int \text{d}\omega\, \frac{\text{d}\phi}{\text{d}\omega} \eta(\omega)\;,
\end{align}
where $A=499\,$cm$^2$, $\eta(\omega)$ is the quantum efficiency, the flux is given by eq.~\eqref{eq:eff_flux}, and the integration range is restricted to the kinematic cuts.

We also show this hypothetical reach in figure~\ref{fig:bounds} as a red dashed line. The good sensitivity of SHIPS, combined with the advantageous energy range, leads to a stronger potential reach compared to the previous probes. However, restricting the energy range also limits the mass range, as can be seen from the sharp cutoff at $m\sim 0.03\,$eV. We stress once again that this is not the true sensitivity of SHIPS to longitudinal DPs but rather the projection of a SHIPS-like grazing incidence helioscope.

\subsection{The new contender: IAXO}\label{ssec:iaxo}

Finally, let us have a look at IAXO (International AXion Observatory), optimistically assuming that it will be built with (or eventually reach) its maximum sensitivity. We will follow the concept proposed in ref.~\cite{Armengaud:2014gea}. In particular, we can see that each of the eight telescopes alone can have an effective area of several thousand cm$^2$ (their figure~11). Especially in the low energy region, which is of highest interest to us, the throughput is at the $80\%$ level and also only mildly dependent on the focal length. We will extract the 5\,m focal length line, knowing that most of the sensitivity will lie at low energies, as in the previous sections.

\subsubsection{BabyIAXO}

Before treating the full IAXO proposal, we will consider the planned predecessor BabyIAXO \cite{IAXO:2020wwp}. We will assume the same optics as in the baseline IAXO proposal and plug a typical angle $\alpha=9\cdot 10^{-3}$ into eq.~\eqref{eq:prob_ex}, given the range of incidence angles $\alpha \in [2.63,15]\cdot 10^{-3}$ (table~4 of ref.~\cite{IAXO:2020wwp}). For the kinematic cutoff, we choose a value close to minimal grazing incidence, $\alpha_\text{cutoff}=3.5\cdot 10^{-3}$, inspired by our findings in appendix~\ref{app:raytracing}.\footnote{Note that BabyIAXO will deploy different optics from the original IAXO proposal \cite{IAXO:2020wwp, Yu:2026ykr, Woo:2026ezp}. We will postpone a detailed analysis of the instrument to future work.} Considering the values shown in their table~1, we rescale the throughput extracted from ref.~\cite{Armengaud:2014gea} by a factor of $0.35\times 0.7/(0.7\times0.8)\sim 0.44$. For the expected sensitivity, we use $10^{-7}\,$keV$^{-1}$s$^{-1}$cm$^{-2}$ from the same table, which also informs us about how the converted flux is focused onto the detector (see also appendix C of ref.~\cite{OShea:2023gqn}): we find an enhancement of $0.77\,$m$^2/(2\times 0.3\,$cm$^2)\simeq 1.3\cdot 10^4$ from the focusing of the geometric area onto the small detector area.
We will assume that this sensitivity can be kept down to energies of $10\,$eV (assuming that the throughput remains at the 80\% level, inspired by ref.~\cite{OShea:2023gqn}).

We show the ensuing sensitivity in figure~\ref{fig:bounds} as a solid dark green line. The upgrade from CAST to BabyIAXO is significant, increasing the sensitivity by several orders of magnitude. To illustrate the impact of the energy threshold, the lighter green line shows the BabyIAXO reach if the detection efficiency can only be maintained down to $100\,$eV. Since both the flux and the conversion probability increase towards lower energies, the reach weakens by a factor of $\sim8$ in $\epsilon$ for $m\lesssim0.03\,$eV; the difference decreases towards larger masses and vanishes for $m\gtrsim0.35\,$eV, where the kinematic cut already restricts the analysis to $\omega>100\,$eV. Maintaining sensitivity at low energies is thus of utmost importance to maximise the reach of helioscopes for longitudinal DPs.

\subsubsection{IAXO}

For the full IAXO, we assume that the instrument can operate at the exposure-limited detection sensitivity of $10^{-8}\,$keV$^{-1}$s$^{-1}$cm$^{-2}$. Focusing the photons onto the small detector area significantly boosts the sensitivity here as well; using the same table as before, we find an enhancement of $2.3\,$m$^2/(8\times 0.15\,$cm$^2)\simeq 1.9\cdot 10^4$ from the ratio of geometric area to detector area. Again, we will restrict the energy range of sensitivity to $[10, 10000]\,$eV. Additionally, to indicate a more speculative, ``ultimate'' limit, we will also consider sensitivity down to energies of $1\,$eV where the flux can be probed fully around its maximum (cf.\ figure~\ref{fig:flux}) and we assume the upgraded configuration parameters (IAXO+) presented in ref.~\cite{IAXO:2019mpb}, in particular the larger geometric area of $3.9\,$m$^2$ and the lower background of $10^{-9}\,$keV$^{-1}$s$^{-1}$cm$^{-2}$. We will further relax the kinematic restrictions generously to $m/\omega<10^{-2}$ assuming optimised optics for this search.

The result is presented in figure~\ref{fig:bounds}. We show the IAXO projection in dark green, where the expected (``ultimate'') sensitivity is presented by a dashed (dotted) line. The expected IAXO sensitivity already clearly outperforms all other bounds, including BabyIAXO, except for the optimistic 2008 reach at masses beyond $\sim 10\,$eV; the small incidence angle of IAXO sets stronger cuts in the energy range and is thus not able to probe this regime. The ``ultimate'' reach is even stronger at all masses because of the lower energy threshold.

One important caveat concerns the signal region: the detector spot of (Baby)IAXO is matched to the image of the inner $\sim0.2\,R_\odot$ of the Sun. Due to the resonance condition (cf.\ section~\ref{sec:solarflux}), the flux at $\omega\sim10\,$eV, which dominates our low-mass sensitivity, originates from a layer at $r\simeq0.8\,R_\odot$ and is thus imaged largely outside this region. This could be addressed by a dedicated longitudinal DP analysis with a signal region adapted to the energy-dependent, ring-like image of the emission layer; such angular information could also provide additional discrimination power against backgrounds \cite{Frerick:2022mjg}. In the following, we nevertheless assume, for simplicity, that the signal is contained within the nominal spot.

\section{Discussion}\label{sec:disc}

\subsection{The novel constraints in the larger picture}

In the following, we will place our new constraints into context. In particular, we should compare them to existing helioscope bounds to see the increase in sensitivity. Furthermore, we ask whether our approach can be used to probe new parameter space, beyond the cooling bound and the absorption in xenon detectors. To this end, we use the compilation of DP bounds from ref.~\cite{AxionLimits} to compare the previous literature with the new results.

For CAST, we found that previous bounds are stronger in the relevant part of parameter space; only for $m\lesssim 3\cdot 10^{-3}\,$eV do our new bounds outperform the literature result.\footnote{This happens approximately when the helioscope becomes too small to contain an oscillation length of the transverse DPs, i.e.\ $m\sim \sqrt{\omega_\text{typ}/L}$ with $L$ the length of the oscillation volume and $\omega_\text{typ}$ a typical DP energy.} For heavier masses up until peak sensitivity at $m\sim 2\,$eV (cf.\ figure~\ref{fig:bounds}), the longitudinal bounds are weaker than the literature results by a factor of 2. In this regime, the low conversion probability counteracts the increased longitudinal flux; we have confirmed that in the limit of a perfect conductor (eq.~\eqref{eq:prob}), the CAST bounds are more than an order of magnitude stronger than previous literature findings.

For the IAXO forecast, on the other hand, even the result with the BabyIAXO configuration is already able to outperform the strongest available bounds from XENON1T \cite{XENON:2021qze} and solar cooling \cite{Li:2023vpv} below $m \simeq 0.05\,$eV, albeit by a small factor $\sim 1.3$ in $\epsilon$. For the baseline IAXO configuration, this already increases to more than a factor of 3. Note also that a factor of $x$ in sensitivity corresponds to a factor of $x^4$ in flux. The optimistic IAXO+ forecast performs even better and would boost the sensitivity by an additional factor of 5 in $\epsilon$. In conclusion, helioscopes could probe more than an order of magnitude of new parameter space in $\epsilon$ for a mass range of $10^{-5}\,\text{eV}\lesssim m\lesssim 0.75\,$eV, where the DarkSRF bounds \cite{Kalia:2025afc} take over at the lower limit and the upper limit is due to the kinematic cuts in our approach and the subsequent drop of sensitivity below other solar DP constraints \cite{An:2020bxd}.

For comparison, we also discuss the results of ref.~\cite{OShea:2023gqn} which investigated IAXO's sensitivity to DPs in great detail. The light blue dotted (dash-dotted) line shows the envelope of the expected sensitivity of IAXO+ (BabyIAXO) to transversely polarised DPs (i.e.\ the standard helioscope setup), longitudinal plasmons that oscillate into transverse DPs due to anisotropies in the solar plasma, and the detection of DPs employing a gas-filled configuration of IAXO, used for resonant DP conversions. The authors find that usually, the strongest sensitivity arises from the standard vacuum setup for transverse DPs.\footnote{The small bump in the BabyIAXO line arises from the resonant conversion in the gas-filled configuration.} This is the reason why the bounds presented in this work are stronger below $m\sim 0.1\,$eV, where the longitudinal DP flux is enhanced, and the kinematic cuts imposed by our framework are not as restrictive. Note that we have just extended the literature sensitivity with a constant scaling towards lower masses. Realistically, the transverse sensitivity of IAXO will suffer from the same suppression as CAST for $m\sim \sqrt{\omega_\text{typ}/L}$.

The grey dashed line shows their speculative IAXO reach assuming that the conversion of the incoming longitudinal DPs could be at the same level as for transverse DPs in the volume of the helioscope \cite{OShea:2023gqn}; even our ``ultimate'' forecast is still far away from this line due to the relatively inefficient conversion on the telescope mirrors. Nevertheless, with this novel method, we have already covered a lot of ground between the grey dashed and light blue dotted lines.

\subsection{A narrow-band boost of the sensitivity}\label{sec:narrowband}

In this section, we will briefly propose another idea that might increase the sensitivity, albeit for a narrow energy range.
To maximise the flux, we consider masses for which the flux peak is still kinematically available, $m\sim 1\,$eV. In this regime, the $m^2$ decoupling of the flux is still mild and the peak of the flux at $\omega\sim 10\,$eV is still kinematically allowed; furthermore, we assume that the reflectors can be constructed from a material close to the perfect conductor limit. This has the consequence that the outgoing angle is expected to be approximately $\beta\sim m/\omega\sim 0.1$, which is relatively large. The aim would now be to design a configuration of mirror and detector that is tuned to the outgoing angle with an incoming grazing angle that is much smaller (as we have seen before, down to $\alpha=2.6\cdot 10^{-3}$). If it is possible to construct such a configuration, the effective conversion factor would not be suppressed. The enhancement comes from the fact that eq.~\eqref{eq:prob} is still $\sim \mathcal{O}(1)\times \epsilon^2$ for the small incoming angle, $m=1\,$eV, and $\omega=10\,$eV. For a conventional telescope, the converted photon will not reach the focus anymore. With our usual kinematic cuts, e.g.\ for $\alpha = 2.6\cdot 10^{-3}$, even the optimistic cuts enforce $\omega \gtrsim 400\,$eV, where the flux is already reduced by 3--4 orders of magnitude compared to the peak at 10\,eV. As the constraints scale with the fourth root of the flux, this could boost the sensitivity by up to an order of magnitude at $m\sim 1\,$eV where the IAXO forecasts already decrease. However, this configuration will be rather specific to a given mass, and thus, it trades the usual broadband sensitivity of the helioscope for a narrow-band improvement. The details will depend on the actual design of the instrument and are beyond the scope of this work. Ideally, one would always remain at $\omega=10\,$eV and then scan over different masses by adjusting the configuration. Such an instrument could come close to reaching the speculative bounds of ref.~\cite{OShea:2023gqn} (cf.\ the dashed grey line in figure~\ref{fig:bounds_full}).

\section{Summary and conclusions}\label{sec:conc}

The dark photon is a highly motivated candidate for new physics, and a broad experimental programme is ongoing to discover it, covering a wide range of masses and coupling strengths. The helioscope is one such experimental approach searching for the solar DP flux, typically by means of vacuum oscillation of \emph{transverse} DPs.

In this paper, we have derived the conversion efficiency of longitudinal DPs on reflectors under grazing incidence in an attempt to make helioscopes sensitive to the longitudinal DP mode as well. We have found that there can be a relevant conversion probability, depending on the reflector material, the (grazing) incoming angle, and the kinematics of the DP. For a fixed incoming angle, the conversion probability increases with decreasing energy, favouring the low energy part of the spectrum. On the other hand, the deviation between incoming and outgoing angles increases as well in this limit, approaching the well-known case of DP dark matter where the outgoing photons propagate along the normal vector of the reflector. Thus, to actually rely on existing and proposed helioscopes, one has to find a compromise between interconversion probability and the requirement for the outgoing photon not to be misaligned too much, which we achieved by introducing kinematic cuts supported by a simple ray tracing simulation.

We have discussed several different helioscope setups: existing results from CAST were used to estimate the sensitivity range of the instrument. The search for so-called chameleon particles turned out to be the most relevant channel for longitudinal DP searches; however, due to the low conversion probability we inferred from the refractive properties of the reflector, the constraints cannot outperform previous literature results. We have further investigated the potential reach of a SHIPS-like setup employing grazing incidence, which could also boost the reach significantly due to its favourable energy range. The key result of this work is the projection for the upcoming IAXO helioscope, which could probe significant parts of unexplored parameter space for $10^{-5}\,$eV$\lesssim m\lesssim 0.75\,$eV. Already the BabyIAXO configuration is competitive with current world-leading bounds and outperforms its forecasted sensitivity in the transverse DP channel, provided that it retains sensitivity down to low energies. We also investigated how an upgraded version (IAXO+) could further improve these results by opening the range of energies and further reducing the background. This configuration would be able to probe deep into unexplored parameter space. We have briefly explored a narrow-band version of a helioscope, tuned to maximise the sensitivity for a chosen mass leveraging the modified outgoing angle.

In summary, the method proposed in this work could make helioscope-driven DP searches competitive with the solar cooling argument and even with absorption searches in liquid xenon detectors. These results also open the door to many new questions; in particular, a dedicated re-analysis of CAST data for this detection mode would be of interest to strengthen the bounds presented here.
Most relevant, however, would be a detailed analysis of IAXO (and BabyIAXO), taking into account all details of the optics to investigate the kinematic cuts in more detail. This would require a meticulous simulation of the DP conversion and subsequent ray tracing, as well as an updated calculation of the angular dependence of the solar DP flux.

\acknowledgments
This work is supported by an ERC StG grant (``AstroDarkLS'', Grant No. 101117510). I thank Amelia Drew, Kai Schmidt-Hoberg, Hyungjin Kim, Martin Mojahed and Felix Kahlh\"{o}fer for discussions and for encouraging me to work on this idea. I am grateful for the very helpful input by Dieter Horns on the functionality of X-ray telescopes and advice on how to model them.

\newpage

\appendix

\section{Modified Poynting's theorem}\label{app:poynting}

This appendix derives the modification of the Poynting vector for a massive vector field.

Let us proceed with the usual approach to Poynting's theorem. However, we should also introduce the Maxwell-Proca equations. We will ignore the kinetic mixing here as we are predominantly interested in the energy stored in DPs. We will also introduce an auxiliary (dark) current $\mathbf{J}^\prime$ and (dark) charge $\rho^\prime$ which are helpful for deriving Poynting's theorem. We then find
\begin{align}
    \nabla\cdot \mathbf{E}^\prime +m^2A_0^\prime&=\rho^\prime\\
    \nabla \times \mathbf{E}^\prime &= -\partial_t \mathbf{B}^\prime\\
    \nabla\cdot \mathbf{B}^\prime &= 0\\
    \nabla \times \mathbf{B}^\prime +m^2 \mathbf{A}^\prime &= \mathbf{J}^\prime +\partial_t\mathbf{E}^\prime\;,
\end{align}
where we have used that $A_0^\prime=g_{\mu 0} (A^0)^\prime=(A^0)^\prime$.
We see clearly that the inhomogeneous (source-driven) equations are modified in a way that reflects that the vector field cannot be completely reduced to electric and magnetic fields, i.e.\ the potentials become physical.

The starting point for Poynting's theorem is the power density delivered to a current (i.e.\ only the electric field contributes here)
\begin{align}
    \mathbf{J}^\prime\cdot \mathbf{E}^\prime&=\left(\nabla \times \mathbf{B}^\prime +m^2 \mathbf{A}^\prime -\partial_t\mathbf{E}^\prime\right)\cdot \mathbf{E}^\prime\\
    &=(\nabla\times \mathbf{E}^\prime)\cdot \mathbf{B}^\prime-\nabla\cdot (\mathbf{E}^\prime\times \mathbf{B}^\prime)+m^2\mathbf{A}^\prime\cdot\mathbf{E}^\prime-\frac{1}{2}\partial_t(\mathbf{E}^\prime)^2\\
    &=-\frac{1}{2}\partial_t\left(\mathbf{E}^{\prime 2}+\mathbf{B}^{\prime 2}\right)- \nabla\cdot (\mathbf{E}^\prime\times \mathbf{B}^\prime)+m^2\mathbf{A}^\prime\cdot (-\nabla A_0^\prime -\partial_t\mathbf{A}^\prime)\\
    &=-\frac{1}{2}\partial_t\left(\mathbf{E}^{\prime 2}+\mathbf{B}^{\prime 2}\right)- \nabla\cdot (\mathbf{E}^\prime\times \mathbf{B}^\prime)-\frac{m^2}{2}\partial_t(\mathbf{A}^{\prime 2})
    -m^2\nabla\cdot (A_0^\prime\mathbf{A}^\prime)+m^2 A_0^\prime \nabla\cdot \mathbf{A}^\prime\\
    &=-\frac{1}{2}\partial_t\left[\mathbf{E}^{\prime 2}+\mathbf{B}^{\prime 2}+m^2\left(A_0^{\prime 2}+\mathbf{A}^{\prime 2}\right)\right]
    - \nabla\cdot (\mathbf{E}^\prime\times \mathbf{B}^\prime+m^2A_0^\prime\mathbf{A}^\prime)\\
    &=-\partial_t u^\prime -\nabla\cdot \mathbf{S}^\prime\;.
\end{align}
Here, we have made use of several vector calculus identities, Maxwell's equations, and the Lorenz condition.
We can identify the first term as a generalised energy density $u^\prime$, now including a mass term. Similarly, the Poynting vector $\mathbf{S}^\prime$ has an additional component, aligned with the vector potential. We see that this describes the propagation of the longitudinal mode: for longitudinal modes, we have seen in eq.~\eqref{eq:b_long} that the magnetic field vanishes and thus the first term of $\mathbf{S}^\prime$ does not carry energy, so that only the potential-dependent term contributes. In the transverse case, the Lorenz condition forces $A_0^\prime \mathbf{A}^\prime=0$ and thus only the ``standard'' Poynting vector contributes.

\section{A more detailed analysis of longitudinal DPs and grazing incidence}\label{app:ter}

In this appendix, we will more carefully derive how grazing incidence reflection works in the setup of DPs. We have seen in the main part of this paper that the masses of interest for this approach are $m\lesssim 10\,$eV to still profit from the large longitudinal flux, whereas heavier DPs are more efficiently probed by their transverse mode. Furthermore, we can read off the plasma frequency of a typical reflector from eq.~\eqref{eq:angle_est} as $\alpha_t=\omega_p/\omega \Rightarrow \omega_p \simeq 88\,$eV for the density of gold ($\rho_\text{Au}\simeq 19\,$g/cm$^{3}$), i.e.\ much higher than the DP masses of interest here. Thus, we can establish $\omega>\omega_p>m$ as a typical hierarchy. However, most of the results discussed in the following are valid outside this hierarchy.

First of all, we need to realise that the vacuum diagonalisation performed as our first step in section~\ref{sec:reflection} is no longer valid in the medium. The coupling to the EM current induces a self-energy contribution mimicking an effective mass to the photon and also off-diagonal terms. Instead of the rotation in eq.~\eqref{eq:rot}, we should perform a different transformation that diagonalises the mass matrix \emph{in the medium}. For simplicity, we will stay at leading order in $\epsilon$. We then find the following transformation for the longitudinal component
\begin{align}\label{eq:rot_mat}
    A_\mu&\to A_\mu+\frac{\epsilon m^2}{\pi_L-m^2} A_\mu^\prime\\
    A^\prime_\mu &\to A^\prime_\mu-\frac{\epsilon \pi_L}{\pi_L-m^2}A_\mu\;.
\end{align}
The rotation is chosen such that the off-diagonal contribution of the mass term and the plasma effects cancel. Note also that it consistently reduces to eq.~\eqref{eq:rot} for $\pi_L\to 0$. Importantly, the DP dispersion relation and the photon plasma dispersion relation are not modified at leading order. The important difference arises in the coupling structure. The effective coupling changes according to $-\epsilon\to \epsilon m^2/(\pi_L-m^2)=-\epsilon /n^2$ where we used $\pi_L=m^2(1-n^2)$ \cite{Raffelt:1996wa}.\footnote{The denominator is what can lead to resonant enhancement of the production e.g.\ in the Sun. In that case, $n^2\simeq 1-\omega_p^2/\omega^2$.} Note that we can assume the same $n^2$ for both transverse and longitudinal modes as the differences are typically suppressed by the low electron velocities in the medium \cite{Braaten:1993jw}.

\begin{figure}
    \centering
    \includegraphics[width=\linewidth]{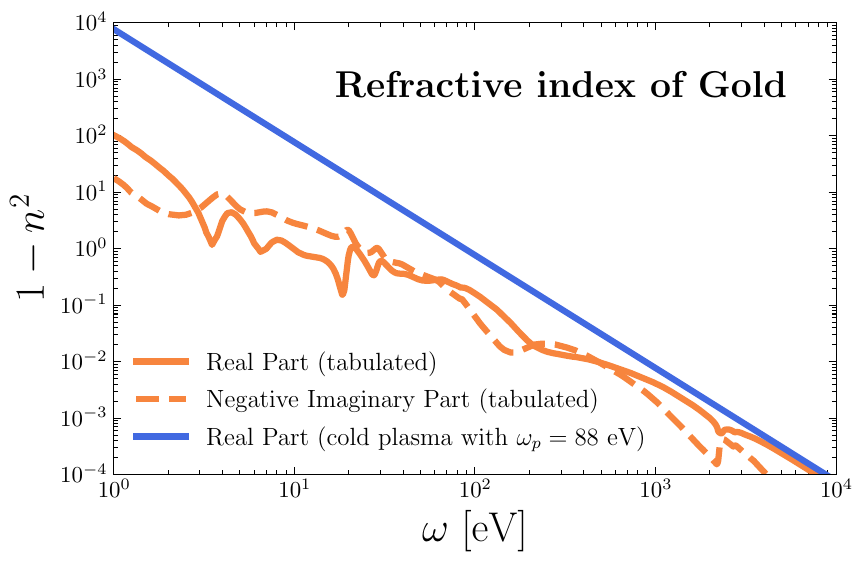}
    \caption{Comparison of the tabulated complex refractive index of gold \cite{10.1063/1.3243762, Polyanskiy2024, HENKE1993181, 10.1117/12.186811} with a simple cold plasma estimate. We flipped the sign of the imaginary part to present it on the same axis.}
    \label{fig:refractive}
\end{figure}

Note that the refractive index will be complex in general. We will use tabulated results to increase the reliability of our results. For simplicity, we will restrict the discussion to the refractive properties of gold; a detailed analysis of the reflective coatings used in the different helioscope searches will be left for future work. For low energies, we use the results from ref.~\cite{10.1063/1.3243762} as presented in ref.~\cite{Polyanskiy2024}, and for higher energies, we rely on the CXRO database \cite{10.1117/12.186811, HENKE1993181}. We stitch together the two data sets at $\omega=55\,$eV with a logarithmic tanh-connection
\begin{align}
    n^2=\frac{n^2_\text{high}+n^2_\text{low}}{2}+\frac{n^2_\text{high}-n^2_\text{low}}{2}\tanh\left(\frac{\log\left(\frac{\omega}{55\,\text{eV}}\right)}{0.1}\right)\;.
\end{align}
The result can be seen in figure~\ref{fig:refractive} where we present both the real and imaginary part of $1-n^2$. We flipped the sign of the imaginary part for better visualisation. We also show the simple cold plasma result $n^2=1-\omega_p^2/\omega^2$ with the aforementioned $\omega_p\simeq 88\,$eV for gold. As can be seen, this is a decent approximation for very energetic photons, but it overestimates the actual value of $|1-n^2|$ for gold in our regime of interest. Furthermore, this simple prescription leads to a purely real $n^2$.

We can now try to match the electric fields together at the boundary. For this, we will write the two propagating modes for each side of the transition (in vacuum and in medium). We note that the two photon-like components have to be transverse. The incoming field is the only part we take as given (cf.\ eq.~\eqref{eq:Einc})
\begin{align}
    \mathbf{E}_\text{vac,in}&= i\begin{bmatrix}
        1\\-\epsilon
    \end{bmatrix} m X\hat{\mathbf{e}}_i e^{-i(\omega t - \mathbf{p}\cdot\mathbf{x})}\\
    \mathbf{E}_\text{vac,out}&= i\begin{bmatrix}
        \epsilon\\1
    \end{bmatrix} m X_\text{vac,out} \hat{\mathbf{e}}_\text{vac,out} e^{-i(\omega t - \mathbf{k}\cdot\mathbf{x})}\\
    \mathbf{E}_{\text{med},m}&= i\begin{bmatrix}
        1\\-\frac{\epsilon}{n^2}
    \end{bmatrix} m X_{\text{med},m}\hat{\mathbf{e}}_{\text{med},m} e^{-i(\omega t - \mathbf{p}_\text{med}\cdot\mathbf{x})}\\    
    \mathbf{E}_{\text{med},n}&= i\begin{bmatrix}
        \mathcal{O}(\epsilon)\\1
    \end{bmatrix} m X_{\text{med},n}\hat{\mathbf{e}}_{\text{med},n} e^{-i(\omega t - \mathbf{k}_\text{med}\cdot\mathbf{x})}\;,
\end{align}
where the labels $m/n$ denote the in-medium propagation eigenstate of the DP-like and photon-like mode, respectively. We use momenta $p$ for the propagation eigenstate with mass $m$ and $k$ for the other one, with subscript $_\text{med}$ to denote the in-medium momenta. Note that we have already denoted the sterile component of the in-medium photon-like wave as $\mathcal{O}(\epsilon)$ to simplify the notation as the exact coupling strength plays no role here.

First, we note that $X_{\text{med},n}\overset{!}{=}\mathcal{O}(\epsilon)$ because otherwise there would be an $\mathcal{O}(\epsilon)$ sterile component propagating in direction $\hat{\mathbf{e}}_{\text{med},n}$ which could not be compensated for by the other in-medium field. The same argument works for $X_\text{vac,out}$. We thus write $X\to\epsilon\tilde{X}$ for the two photon-like modes and neglect their upper components as they are $\mathcal{O}(\epsilon^2)$.

Continuity of the sterile component thus requires $X_{\text{med},m}=X$ and $\hat{\mathbf{e}}_{\text{med},m}=\hat{\mathbf{e}}_i$, which also implies that $\mathbf{p}_\text{med}=\mathbf{p}$ as the leading order DP-like dispersion relations are the same in vacuum and in medium.\footnote{It could be interesting to investigate whether the resonant absorption of incoming DPs might lead to observable effects.} This is all the information that can be drawn from the sterile component.

We will now focus on the interacting component. Due to the modified coupling strength, the transition between vacuum and reflector becomes non-trivial. To allow for a continuous behaviour, all four components must cancel. We will now split the waves into the part propagating parallel and perpendicular to the reflector. We will begin with the parallel component: we label the incoming angle $\alpha$ (as before, measured w.r.t.\ the reflector surface), $\beta$ the outgoing angle, and $\beta_m$ the in-medium angle of the mostly interacting wave; $\alpha_m=\alpha$ as derived in the previous paragraph. Then we find
\begin{align}
    &-i\epsilon m X \cos\alpha e^{-i(\omega t - px_s\cos\alpha)}+i\epsilon m\tilde{X}_{\text{vac,out},\parallel} e^{-i(\omega t - kx_s\cos\beta)}\\
    \overset{!}{=}&-i\frac{\epsilon}{n^2}m X \cos\alpha e^{-i(\omega t - px_s\cos\alpha)}+i\epsilon m\tilde{X}_{\text{med},n,\parallel} e^{-i(\omega t - k_\text{med}x_s\cos\beta_m)}\;.
\end{align}
We can use the exponential factors to realise
\begin{align}
    p\cos\alpha&=k\cos\beta=k_\text{med}\cos\beta_m\\
    \Rightarrow \cos\beta&=\frac{p}{k}\cos\alpha=\sqrt{1-\frac{m^2}{\omega^2}}\cos\alpha\\
    \Rightarrow \cos\beta_m&=\frac{p}{k_\text{med}}\cos\alpha=\frac{\sqrt{1-\frac{m^2}{\omega^2}}}{\sqrt{n^2}}\cos\alpha\;.\label{eq:betam}
\end{align}
The second line is the same result as we found before. The third line is the crucial result: while $\sqrt{1-\frac{m^2}{\omega^2}}<1$, and thus always physically allowed, we notice that in our setup $\sqrt{1-m^2/\omega^2}/|n|>1$ is possible, and for general (complex) $n^2$, the momentum becomes complex.\footnote{Note that the dispersion relation of the \emph{transverse} component is simply $\omega^2-k^2=\pi_T=\omega^2(1-n^2)$, unlike for the longitudinal one.} Assuming for now the simplified relation $n^2=1-\omega_p^2/\omega^2$, we get the critical angle $\alpha_t$ by demanding
\begin{align}
    &\frac{\sqrt{\omega^2-m^2}}{\sqrt{\omega^2-\omega_p^2}}\cos\alpha_t=1\\
    \Rightarrow &1-\frac{\alpha_t^2}{2}\simeq \cos\alpha_t = \frac{\sqrt{\omega^2-\omega_p^2}}{\sqrt{\omega^2-m^2}}\simeq 1- \frac{\omega_p^2-m^2}{2\omega^2}\;,
\end{align}
where we went to the limit $\omega\gg \omega_p$. If the mass is negligible, we have restored the result in eq.~\eqref{eq:angle_est}. Note that with the chosen values for $m$ and $\omega_p$, the relative deviation is $\sqrt{\omega_p^2-m^2}/\omega_p\simeq 0.99$, i.e.\ the deviation is at the percent level, and thus the critical angle remains essentially unaffected.

In this simplified scenario, this implies for the normal component of the in-medium momentum
\begin{align}
    k_\text{med}\sin\beta_m =i\sqrt{\omega_p^2+p^2\cos^2\alpha-\omega^2} \;.
\end{align}
Thus, a so-called evanescent wave, rapidly damped in the normal direction, propagates along the surface of the reflector. For this argument to work, the material needs to have a sufficient thickness to completely damp the wave. However, as the wave is very photon-like, and the telescope was designed to reflect photons, this assumption is well justified. To quantify this, for sufficiently low angles, the damping scale can approximately be taken as $\omega_p\sim 88\,$eV which sets a length scale of $\sim 2\,$nm which has to be compared with the 250\,nm given as an example from XMM in ref.~\cite{IAXO:2020wwp}.

For general complex $n$, we find
\begin{align}
     k_\text{med}\sin\beta_m =\omega\sqrt{n^2-\cos^2\alpha\left(1-\frac{m^2}{\omega^2}\right)} \;.
\end{align}
The interpretation is straightforward: for normal incidence ($\cos\alpha=0$), the damping of the in-medium photon-like mode is set by $\text{Im}(n)$. For grazing incidence, there is an additional suppression in the direction normal to the surface due to total external reflection if $\alpha$ falls below the critical angle $\alpha_t$ (which requires $\text{Re}(n)<1$); this scenario is of particular relevance if $\text{Im}(n)\ll 1$. In general, it is important to choose the branch of $\sqrt{n^2}$ that leads to damping of the wave propagating into the medium.

Now that the exponentials are fixed, we can consider the amplitudes. Including the Lorenz conditions, we find
\begin{align}
 \tilde{X}_{\text{vac,out},\parallel} 
   &=\left(-\frac{1-n^2}{n^2}\right) X \cos\alpha +\tilde{X}_{\text{med},n,\parallel}\\
   \tilde{X}_{\text{vac,out},\perp} 
   &=\left(1-\frac{n^2}{n^2}\right) X \sin\alpha +n^2\tilde{X}_{\text{med},n,\perp}
   =n^2\tilde{X}_{\text{med},n,\perp}\\
   \sin\beta\tilde{X}_{\text{vac,out},\perp} &=\cos\beta \tilde{X}_{\text{vac,out},\parallel}\\
   \sin\beta_m\tilde{X}_{\text{med},n,\perp} &=-\cos\beta_m \tilde{X}_{\text{med},n,\parallel}\;,  
\end{align}
where the second line takes into account that not the normal component of $\mathbf{E}$ but of $\mathbf{D}=n^2\mathbf{E}$ has to be continuous through the surface.
After a bit of work, we find the expression
\begin{align}
    \tilde{X}_{\text{vac,out},\parallel} &=(n^2-1)X\cos\alpha\left[n^2+\frac{\tan\beta_m}{\tan\beta}\right]^{-1}\;.
\end{align}
For the normal component, we then find
\begin{align}
    \tilde{X}_{\text{vac,out},\perp}&=\frac{\tilde{X}_{\text{vac,out},\parallel}}{\tan\beta}\\
    &=(n^2-1)X\cos\alpha\left[n^2\tan\beta+\tan\beta_m\right]^{-1}\;.
\end{align}
To arrive at the outgoing Poynting vector, we need the magnetic field as well. We write
\begin{align}
    \langle \mathbf{S}\rangle &= \frac{1}{2}\text{Re}\left(\mathbf{E}_\text{vac,out}\times \mathbf{B}_\text{vac,out}^\ast\right)\\
    &=\frac{1}{2\omega}\text{Re}\left(\mathbf{E}_\text{vac,out}\times \left(\mathbf{k}\times\mathbf{E}_\text{vac,out}^\ast\right)\right)\\
    &=\frac{1}{2\omega}|\mathbf{E}_\text{vac,out}|^2\mathbf{k}=\frac{1}{2}|\mathbf{E}_\text{vac,out}|^2\hat{\mathbf{e}}_k\\
    \Rightarrow |\langle \mathbf{S}\rangle| &=\frac{\epsilon^2 m^2 }{2}\left[|\tilde{X}_{\text{vac,out},\perp}|^2+|\tilde{X}_{\text{vac,out},\parallel}|^2\right]\\
&=\frac{\epsilon^2 m^2 X^2\cos^2\alpha}{2}\left|1-n^2\right|^2\frac{1+\frac{1}{\tan^2\beta}}{\left|n^2+\frac{\tan\beta_m}{\tan\beta}\right|^2}\\
    &\xrightarrow[]{|n^2|\to \infty}\frac{\frac{1}{2}\epsilon^2m^2\omega^2X^2\cos^2\alpha}{\omega^2\sin^2\alpha+m^2\cos^2\alpha}\;.
\end{align}
As alluded to in the main text, we see that the signal for total external reflection of DPs receives an additional suppression from the fact that $|n|\sim 1$ for a variety of energies (cf.\ figure~\ref{fig:refractive}). Note further that the limit of a perfect conductor ($|n^2|\to \infty$ at fixed angle) restores the result of section~\ref{sec:reflection}. 
The effective conversion probability is then given by
\begin{align}
    \mathcal{P}=\frac{ |\langle\mathbf{S}\rangle|}{|\langle\mathbf{S}^\prime\rangle|}
    &=\epsilon^2 \frac{m^2\cos^2\alpha}{(\omega^2\sin^2\alpha+m^2\cos^2\alpha)\sqrt{1-\frac{m^2}{\omega^2}}}\frac{\left|1-n^2\right|^2}{\left|n^2+\frac{\tan\beta_m}{\tan\beta}\right|^2}=\mathcal{P}_0\frac{\left|1-n^2\right|^2}{\left|n^2+\frac{\tan\beta_m}{\tan\beta}\right|^2}\;,\label{eq:prob_ter}
\end{align}
with $\mathcal{P}_0$ as in eq.~\eqref{eq:prob}.
The factorised form of eq.~\eqref{eq:prob_ter} makes the connection to the perfect conductor of section~\ref{sec:reflection} explicit. Using $\tan\beta_m/\tan\beta = k_\text{med}\sin\beta_m/(\omega\sin\beta)$ and expanding for $|n|\gg 1$, where $k_\text{med}\sin\beta_m=\omega\sqrt{n^2-\cos^2\alpha\,(1-m^2/\omega^2)}\simeq n\,\omega$ (with $\text{Im}\,n\geq 0$), we find
\begin{align}\label{eq:pc_limit}
    \frac{\mathcal{P}}{\mathcal{P}_0}\simeq\frac{\left|1-\frac{1}{n^{2}}\right|^2}{\left|1+\frac{1}{n\sin\beta}\right|^2}\;.
\end{align}
The perfect conductor result $\mathcal{P}\to\mathcal{P}_0$ is therefore only recovered if
\begin{align}
    |n|\gg\frac{1}{\sin\beta}=\frac{\omega}{\sqrt{\omega^2\sin^2\alpha+m^2\cos^2\alpha}}\simeq
    \begin{cases}
        1/\alpha\;,&\quad \omega\alpha\gg m\\
        \omega/m\;,&\quad \omega\alpha\ll m
    \end{cases}\;,
\end{align}
which is much stronger than the naive requirement $|n|\gg 1$. Interestingly, grazing incidence thus actually prevents the reflector from being a good converter: the reflected photon has a normal wavenumber $\omega\sin\beta$ which is small compared to $\omega$, and the reflector only acts as a perfect conductor if the field inside the material varies on much shorter scales, $|k_\text{med}\sin\beta_m|\gg\omega\sin\beta$. For a lossless conductor ($n^2<0$, i.e.\ $n$ purely imaginary), eq.~\eqref{eq:pc_limit} becomes $\mathcal{P}/\mathcal{P}_0\simeq\left[1+1/(|n|^2\sin^2\beta)\right]^{-1}$, whereas absorption ($\text{Re}\,n>0$) already introduces a correction at first order, $\mathcal{P}/\mathcal{P}_0\simeq 1-2\,\text{Re}\left[(n\sin\beta)^{-1}\right]$.

Let us conclude this appendix with a final observation: the angle of total external reflection is not a hard threshold for the conversion of DPs on the reflector; however, for larger values of $\alpha$ the medium can support the propagation of a mostly interacting component which in turn means that the outgoing flux will be even smaller. Thus, grazing incidence will tend to increase the conversion probability in the regime where the real part of the refractive index is very close to unity and the imaginary part is negligible, as is typical in the X-ray range. On the other hand, grazing incidence prevents a good conductor from becoming a perfect conductor due to the condition $|n|\gg1/\sin\beta\gg1$. However, for large incoming angles $\alpha$, the perfect conductor conversion probability suffers from an $m^2/\omega^2$ suppression as well. Overall, this variety of different effects might allow one to construct an optimised configuration.

\section{Ray tracing of converted photons in a toy Wolter-I telescope}\label{app:raytracing}

In the main text we used a simple analytical argument to estimate the kinematic cutoff: a photon produced by a DP conversion leaves the mirror at a slightly larger angle than a specularly reflected photon. Beyond some value of $m/\omega$, it therefore no longer reaches the detector. In this appendix we check this estimate with a geometric three-dimensional ray tracer for a nested Wolter-I telescope with IAXO-like optics. The tracer only follows the geometry of the converted photons. It answers which fraction of the converted photons is focused onto the detector, as a function of $m/\omega$. The conversion probability itself is applied separately on the analytical side; the tracer only records the grazing angle of each conversion, so that it can be weighted later.

Throughout this appendix the optical axis is the $z$ axis. The focal point is at the origin, and the Sun is in the direction $+z$. The junction between the two mirrors of a shell (the intersection plane) is at the focal length $z=F$. 

\subsection{Telescope geometry}\label{app:rt_geometry}

A Wolter-I telescope consists of nested shells, each consisting of a paraboloid (the primary mirror, which the incoming light hits first) followed by a hyperboloid (the secondary mirror). The far focus of the hyperboloid coincides with the focus of the paraboloid; two grazing-incidence reflections bring rays parallel to the axis to the near focus of the hyperboloid \cite{B_Aschenbach_1985}. In the simulation, every shell is described by an exact quadric surface. A shell is fixed by its radius $r_0$ at the intersection plane $z=F$. We assume an equal-angle design: a ray parallel to the axis hits both mirrors at the same grazing angle $\alpha$ when they meet at the intersection plane. A ray parallel to the axis that is reflected at the junction on the primary, whose slope there is $\tan\alpha$, leaves at an angle $2\alpha$ to the axis. If the secondary, with slope $\tan 3\alpha$ at the junction, is hit under the same grazing angle $\alpha$, the ray is deflected by another $2\alpha$ and travels from $(r_0,F)$ to the focus at an angle $4\alpha$, which gives
\begin{align}
    \tan 4\alpha = \frac{r_0}{F}\;.
\end{align}
The primary is the paraboloid
\begin{align}
    r^2 &= x^2+y^2 = 4f_p\,(z-z_v)\;,\\
    f_p&=\frac{r_0\tan\alpha}{2}\;,\qquad z_v = F-\frac{r_0^2}{4f_p}\;,
\end{align}
whose slope at $(r_0,F)$ is $\mathrm{d}r/\mathrm{d}z=\tan\alpha$. Its focus lies at $z_p=z_v+f_p$. The secondary is the hyperboloid with foci at $z_p$ and at the origin which passes through $(r_0,F)$. With $d_1=\sqrt{r_0^2+F^2}$ and $d_2=\sqrt{r_0^2+(F-z_p)^2}$, the distances of that point to the two foci, we find
\begin{align}
    r^2 &= \frac{b^2}{a^2}\,(z-z_c)^2 - b^2\;,\\
    a&=\frac{d_2-d_1}{2}\;,\qquad z_c=\frac{z_p}{2}\;,\qquad b^2=z_c^2-a^2\;.
\end{align}
Both surfaces are thus of the form
\begin{align}\label{eq:rt_quadric}
    Q(x,y,z)\equiv x^2+y^2-Az^2-Bz-C=0\;,
\end{align}
with $A=0$ for the paraboloid. The primary is restricted to $F\le z\le F+L_p$ and the secondary to $F-L_h\le z\le F$, where $L_{p/h}$ is the length of the paraboloid/hyperboloid mirror segments.

The parameters follow the IAXO conceptual design \cite{Armengaud:2014gea}: focal length $F=5\,$m, 123 shells, and grazing angles from $2.63$ to $15.0\,$mrad.\footnote{It is possible that these grazing angles are not defined at the intersection but at the beginning of the primary; for our toy model, we will stick with our choice and postpone an analysis of the detailed (Baby)IAXO optics to future work.} Through $\tan4\alpha=r_0/F$, the grazing angles fix the radii of the innermost and outermost shell at the intersection plane to $52.6\,$mm and $300.4\,$mm in good agreement with the literature values of $50\,$mm and $300\,$mm. Ref.~\cite{Armengaud:2014gea} gives neither the segment length nor the substrate thickness. We take $L_p=L_h=L=225\,$mm and a glass thickness of $0.21\,$mm, the values of the NuSTAR-type segments that are foreseen for the inner part of the BabyIAXO optic \cite{Yu:2026ykr}.

The radii of the intermediate shells are not listed in ref.~\cite{Armengaud:2014gea}. We therefore pack the shells as densely as possible without obstructing the aperture for on-axis rays. Consecutive shells are separated by a constant gap $g$ beyond the entrance edge (including the glass thickness $t$) of the inner one:
\begin{align}
    r_{0,i+1} \simeq r_{0,i} + L\tan\alpha_i + t + g\;,
\end{align}
where $L\tan\alpha_i$ is the approximate width that is required for parallel rays to be able to reach the junction point.
The gap $g$ is chosen numerically such that the 123rd shell ends at the outermost radius, which gives $g\simeq0.15\,$mm.

\subsection{Rays from the solar core}\label{app:rt_source}

The Sun is so far away that every point of its disk illuminates the telescope with a plane wave. A ray is therefore characterised by two independent quantities: its direction is set by the point on the solar disk it comes from, and its starting point is set by where it enters the aperture.
The starting points are drawn uniformly in area over the entrance annulus, $r_\text{in}\le r\le r_\text{out}$, at the height $z_\text{start}=F+L+1\,$mm where we introduced the small offset for technical reasons. With $U'$ uniform in $[0,1]$ and $\phi$ uniform in $[0,2\pi)$,
\begin{align}
    r=\sqrt{r_\text{in}^2+U'\left(r_\text{out}^2-r_\text{in}^2\right)}\;,\qquad \mathbf{o}=(r\cos\phi,\;r\sin\phi,\;z_\text{start})\;,
\end{align}
where $r_\text{in}=52.6\,$mm is the intersection radius of the innermost shell, and $r_\text{out}=303.9\,$mm is the outer edge of the outermost primary, including its glass thickness. 

We assume that the DPs come from the same region as the solar axions: a uniformly bright disk of angular radius $\theta_\text{src}=3'$ ($0.87\,$mrad or $0.2\,R_\odot$), centred on the optical axis. The direction of each ray is drawn uniformly over this disk. With $U$ uniform in $[0,1]$ and $\psi$ uniform in $[0,2\pi)$,
\begin{align}
    \theta=\theta_\text{src}\sqrt{U}\;,\qquad \hat{\mathbf{d}}=(\sin\theta\cos\psi,\;\sin\theta\sin\psi,\;-\cos\theta)\;,
\end{align}
where the square root ensures a uniform density per solid angle for $\theta_\text{src}\ll1$. An image of this disk in the focal plane has the radius $F\theta_\text{src}\simeq 4.4\,$mm.

\subsection{Propagation, reflection and conversion}\label{app:rt_tracing}

A ray can be described by $\mathbf{P}(t)=\mathbf{o}+t\,\hat{\mathbf{d}}$, where $t$ is the path length. Inserting it into eq.~\eqref{eq:rt_quadric} gives a quadratic equation in $t$ for every surface. The relevant crossing is the smallest root with $t>0$ that falls into the $z$ range of the mirror segment; the quadric itself continues beyond the glass. Given this crossing, the normal vector pointing towards the reflective side of the surface is
\begin{align}
    \hat{\mathbf{n}}\propto -\nabla Q = 2\left(-x,\;-y,\;Az+\tfrac{B}{2}\right)\;.
\end{align}

\paragraph{Conversion and reflection.} Both processes follow from conserving the momentum parallel to the surface. An incoming particle with direction $\hat{\mathbf{d}}$ and momentum $|\mathbf{p}|=\sqrt{\omega^2-m^2}$ produces a photon with $|\mathbf{k}|=\omega$. The photon direction is
\begin{align}
    \hat{\mathbf{k}} &= \sqrt{1-\frac{m^2}{\omega^2}}\;\hat{\mathbf{d}}_\parallel + k_n\,\hat{\mathbf{n}}\;,\label{eq:rt_kout}\\
    k_n&=\sqrt{1-\left(1-\frac{m^2}{\omega^2}\right)|\hat{\mathbf{d}}_\parallel|^2}\;,
\end{align}
with the component of the incoming direction parallel to the surface, $\hat{\mathbf{d}}_\parallel=\hat{\mathbf{d}}-(\hat{\mathbf{d}}\cdot\hat{\mathbf{n}})\hat{\mathbf{n}}$. The first term conserves the parallel momentum; the second fixes $|\hat{\mathbf{k}}|=1$ and points away from the surface.
With the grazing angles $\sin\alpha=-\hat{\mathbf{d}}\cdot\hat{\mathbf{n}}$ and $\sin\beta=\hat{\mathbf{k}}\cdot\hat{\mathbf{n}}$, this is equivalent to $\cos\beta=\sqrt{1-m^2/\omega^2}\cos\alpha$, cf.\ eq.~\eqref{eq:angle}. For small angles,
\begin{align}
    \beta\simeq\sqrt{\alpha^2+\frac{m^2}{\omega^2}}\;\xrightarrow{m/\omega\ll\alpha}\;\alpha+\frac{m^2}{2\omega^2\alpha}\;.
\end{align}
For $m=0$, eq.~\eqref{eq:rt_kout} is ordinary specular reflection, which we use for every reflection of a photon. The outgoing direction therefore depends on $m$ and $\omega$ only through the ratio $m/\omega$, so that all results below can be read as functions of $m/\omega$ and later folded with the spectrum.

\paragraph{Photon tracing.} A photon is propagated from surface to surface. At each step we determine the nearest crossing with any of the 246 mirror surfaces. If the photon reaches the surface from the coated (i.e.\ reflecting) side ($\hat{\mathbf{d}}\cdot\hat{\mathbf{n}}<0$), it is specularly reflected with reflectivity one. If it reaches it from the uncoated back side, we assume it to be lost. After at most four interactions, photons that move towards the focal plane are propagated to $z=0$, and their position there is recorded. In the final result, however, we only take into account those photons that were created at the primary and reflected at the secondary mirror.\footnote{If we only propagate photons, we have to replace the creation by the reflection on the primary mirror.}

The glass thickness is taken into account only at the end faces of the shells: we check whether photons leaving the optics hit the rear end faces. Furthermore, if we track ordinary photons, we discard those that pass through the finite width of the mirrors at the aperture. These block $8.7\,\%$ of the entrance annulus.

\paragraph{Dark photons.} DPs interact only through the $\epsilon$-suppressed coupling. We therefore let them propagate in straight lines through all shells without being blocked or attenuated. For every DP ray we determine its crossings with all primary mirrors. We assume that only crossings from the coated side produce a photon, with one photon per crossing and the direction given by eq.~\eqref{eq:rt_kout}.\footnote{The interface conditions will also give rise to an ordinary photon wave if the DP passes through a mirror from the back.} The photon is then traced as described above. We count it as \emph{focused} if it is reflected exactly once on a secondary and reaches the focal plane. The source radius is smaller than the smallest grazing angle, $\theta_\text{src}=0.87\,\text{mrad}<\alpha_\text{min}=2.63\,$mrad. A DP can therefore cross at most one primary from its coated side; we checked this explicitly. For $m/\omega\to0$, the converted photons are identical to specularly reflected photons.

\paragraph{Signal efficiency.} As the signal region we take the detector entrance window, a disk of radius $R_\text{w}=8.5\,$mm (1.7\,cm diameter) centred on the optical axis in the focal plane as indicated in figure~33 of ref.~\cite{IAXO:2020wwp}. The geometric signal efficiency is
\begin{align}\label{eq:rt_eff}
    \eta_r(m/\omega)=\frac{N_r(m/\omega)}{N_r(m/\omega\to0)}\;,
\end{align}
where $N_r$ is the number of converted photons that arrive inside a circular spot with radius $r$ for the same set of DP rays. Each conversion is counted with equal weight.

For orientation, we also trace ordinary X-rays from an on-axis point source with unit reflectivity. Of the entrance annulus, $84.5\,\%$ is focused with two reflections, i.e.\ a geometric effective area of $\simeq2380\,$cm$^2$ for a single telescope module. The remaining rays are blocked by the front faces or pass between the shells without the correct pair of reflections.

\subsection{Results}\label{app:rt_results}

\begin{figure}[t]
    \centering
    \includegraphics[width=\linewidth]{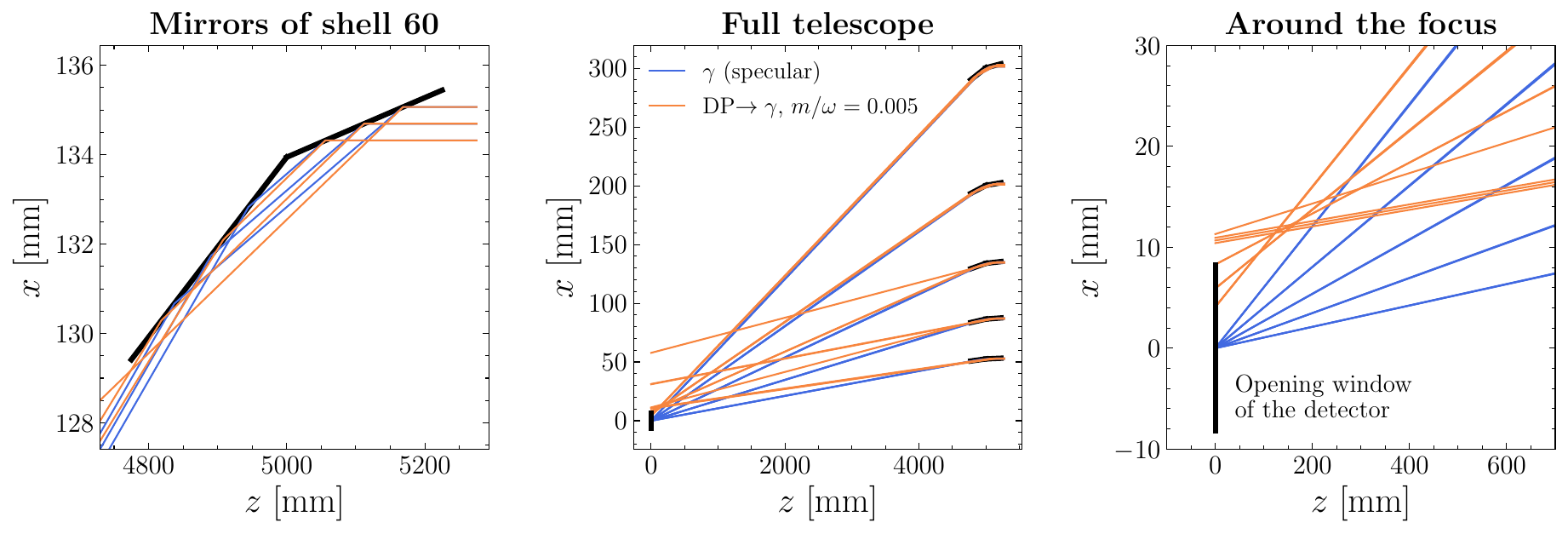}
    \caption{Cut through the simulated IAXO-like telescope in the $z$-$x$-plane. Black lines are mirror segments (primaries for $z>F=5000\,$mm, secondaries for $z<F$). Blue lines are photons from an on-axis source, reflected specularly on a primary and a secondary. Orange lines are DPs that propagate parallel to the axis, convert on a primary, and whose photons are traced onwards. We use $m/\omega=5\cdot10^{-3}$. \textbf{Left}: zoom on shell 60 ($r_0=134\,$mm, $\alpha=6.7\,$mrad). \textbf{Centre}: shells 0, 30, 60, 90 and 122, from the entrance to the focal plane. (Note the indexing beginning from 0.) \textbf{Right}: the region around the focus with the detector entrance window ($|x|<8.5\,$mm, thick black line at $z=0$).}
    \label{fig:rt_paths}
\end{figure}

Figure~\ref{fig:rt_paths} shows the geometry of a few shells and illustrates the mechanism; we generate incoming rays exactly parallel to the optical axis. The specularly reflected photons (blue) all pass through the focus. The converted photons (orange) leave the primary at the larger angle $\beta>\alpha$. They therefore hit the secondary further downstream, under a grazing angle that differs from $\alpha$ by $\beta-\alpha$. After the second reflection, they travel towards the focal plane at an angle that is shallower by about $\beta-\alpha$ than that of a specular ray. In the focal plane, this displaces them from the focus by roughly $F(\beta-\alpha)\simeq F\,m^2/(2\omega^2\alpha)$. The displacement is largest for the inner shells, which have the smallest grazing angles. For a photon converted close to the entrance end of a primary, the steeper path can even miss the secondary completely: the photon leaves the telescope through the exit plane without a second reflection and ends up far from the focus. In the left panel this is visible for the outermost converted photon, which travels on a straight line below the secondary.

\begin{figure}[t]
    \centering
    \includegraphics[width=\linewidth]{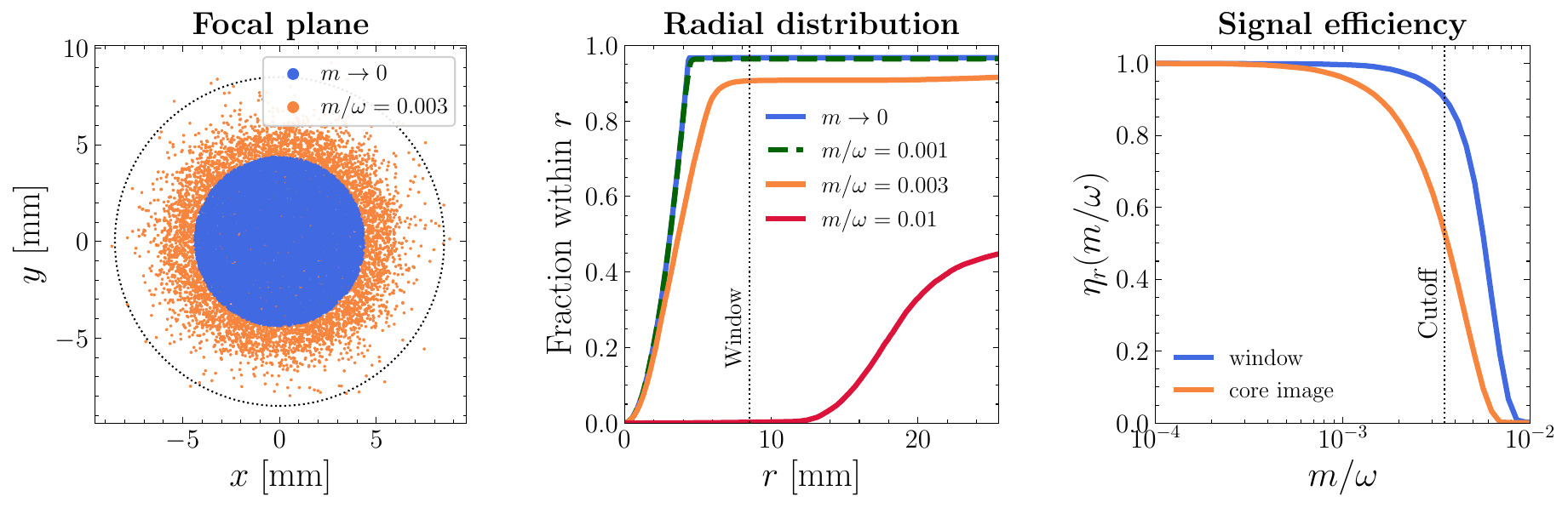}
    \caption{Converted photons from DPs emitted by the solar core (uniform disk of radius $3'$). \textbf{Left}: positions in the focal plane of the focused photons for $m/\omega\to0$ (blue) and $m/\omega=3\cdot10^{-3}$ (orange). The dotted circle shows the detector entrance window. \textbf{Centre}: fraction of all converted photons that arrive within a radius $r$ of the optical axis, for several values of $m/\omega$. The dotted line marks the window radius. \textbf{Right}: geometric signal efficiency $\eta_r$, eq.~\eqref{eq:rt_eff}, i.e.\ the number of converted photons inside the window (blue) or the solar core image (orange) relative to the case $m/\omega\to0$. The dotted line marks the kinematic cutoff adopted in the analysis, $m/\omega=3.5\cdot10^{-3}$.}
    \label{fig:rt_core}
\end{figure}

Figure~\ref{fig:rt_core} shows the result for DPs from the solar core. We generate $2\cdot 10^4$ rays.

\paragraph{Left panel.} For $m/\omega\to0$, the focused photons form the image of the uniformly bright core, a disk of radius $\simeq4.4\,$mm. The image is sharp because the simulated mirrors are ideal. At $m/\omega=3\cdot10^{-3}$, the image is smeared out to radii of several mm. The displacement in the focal plane is radial, so the image stays rotationally symmetric. The photons converted on the inner shells, with $\alpha$ comparable to $m/\omega$, are displaced the most. For this mass, practically all focused photons still arrive within the window. However, the larger area of the target region needs to be taken into account; alternatively, we might only want to account for the fraction of events that is still contained within the original image of the Sun.

\paragraph{Central panel.} This panel shows the cumulative radial distribution of all converted photons, normalised to the number of conversions.
\begin{itemize}
    \item For $m/\omega\to0$ and $10^{-3}$, essentially all photons ($\simeq97\,\%$) arrive within the image of the core.
    \item The curve for $m/\omega=3\cdot10^{-3}$ rises more slowly and saturates at $\simeq91\,\%$ inside the window. The missing $\simeq9\,\%$ are photons that missed the secondary and arrive far outside the plotted range. 
    \item For $m/\omega=10^{-2}$, only $43\,\%$ of the converted photons are still focused, i.e.\ reflected exactly once on a secondary. Those photons that are still focused lie outside the detector window. Another $35\,\%$ miss the secondary completely and reach the focal plane far from the axis. The remaining $22\,\%$ are absorbed on the back side of a neighbouring shell or at the rear edge of a secondary.
\end{itemize}

\paragraph{Right panel.} The efficiency $\eta_r$ is shown for $r=8.5\,$mm and $r=4.4\,$mm. The former accounts for the overall efficiency of measuring converted DPs with a given helioscope using the maximum spot size available. The latter indicates how strongly the original image of the solar core is distorted. The following points are valid for $r=8.5\,$mm, but the results are qualitatively similar for $r=4.4\,$mm:
\begin{itemize}
    \item Up to $m/\omega\simeq5\cdot10^{-3}$: the loss comes almost entirely from photons that miss the secondary. These are photons converted near the entrance of the primaries, especially in the innermost shells. 
    \item Beyond $m/\omega\simeq5\cdot10^{-3}$: the displacement $F(\beta-\alpha)$ exceeds the chosen radius for an increasing number of shells, and the efficiency drops steeply. Requiring $F\,m^2/(2\omega^2\alpha)\lesssim r$ gives $m/\omega\lesssim\sqrt{2\alpha r/F}\simeq3\cdot10^{-3}\sqrt{r/(8.5\,\text{mm})}$ ($\simeq7\cdot10^{-3}\sqrt{r/(8.5\,\text{mm})}$) for the innermost (outermost) shell, in agreement with the simulation.
\end{itemize}
The image of the solar core gets depopulated before the detector turns fully blind, as is clear from the offset between the orange and blue curves. At the adopted cutoff $m/\omega=3.5\cdot 10^{-3}$, the efficiency is $\eta_r\simeq0.5$ in the core region. The sharp cutoff used in the main text is therefore a good description of the IAXO optics with this signal region.

\subsection{Approximations and assumptions}\label{app:rt_assumptions}

The results above rely on the following simplifications.

\begin{itemize}
    \item \textbf{Geometry only.} Each conversion is counted with equal weight, and $\eta_r$ is a purely geometric acceptance. In reality each photon carries the conversion probability of eq.~\eqref{eq:prob_ex}, which depends on the local grazing angle. 
    \item \textbf{Ideal mirrors.} The mirrors are exact quadrics without figure errors, micro-roughness or misalignment. The secondary reflection has unit reflectivity, independent of energy, angle and polarisation. The converted photons are polarised, and the reflectivity at the second mirror will in general differ from that for unpolarised X-rays. Real optics have a finite point spread function. For IAXO, ref.~\cite{Armengaud:2014gea} assumes a half-power diameter of $1'$ and an $80\,\%$ encircled-energy diameter of $2'$ ($0.58\,$mrad, i.e.\ $\simeq3\,$mm in the focal plane). This blurring is not included; it would smear the images in figure~\ref{fig:rt_core}.
    \item \textbf{Telescope design.} We assume an idealised IAXO module. Focal length, number of shells and the range of grazing angles are taken from ref.~\cite{Armengaud:2014gea}. The segment length, substrate thickness and the radial packing of the shells are our own choices, inspired by ref.~\cite{Yu:2026ykr}. Following from the quoted grazing angles, the innermost radius is $52.6\,$mm rather than the $50\,$mm quoted in ref.~\cite{Armengaud:2014gea}. The shells are complete surfaces of revolution: segmentation and the support structure are ignored. They would reduce the throughput by an amount that is the same for photons and converted DPs, so they cancel in $\eta$ to first approximation. The glass thickness is only taken into account at the end faces of the shells. We simulate a single telescope module. IAXO foresees eight identical modules, and BabyIAXO uses different optics \cite{IAXO:2020wwp, Yu:2026ykr}.
    \item \textbf{Conversion only on the primaries.} Only conversions on the coated side of the primary mirrors are simulated. DPs also cross the secondaries, but photons produced there are not focused by a second reflection. They would at most add a diffuse contribution in the focal plane. Conversions on the uncoated back sides of the shells and on the end faces are neglected.
    \item \textbf{Source model \& region.} We assume that the DP signal comes from the same region as the solar axions: a uniformly bright, on-axis disk of $3'$ radius, as in ref.~\cite{Yu:2026ykr}. The sensitivity estimates in the main text normalise the background to a much smaller spot area of $\simeq0.15$--$0.3\,$cm$^2$ per telescope \cite{Armengaud:2014gea, IAXO:2020wwp}. For a precise prediction, we would need the angular flux of the longitudinal DP (see refs.~\cite{Redondo:2015iea, Frerick:2022mjg} for the transverse component). We further assume perfect pointing and tracking.
     We count the photons with a detection efficiency that is uniform over the window.
    \item \textbf{Monte Carlo statistics.} The numbers quoted in the text and the figures use $2\cdot10^4$ rays. The statistical uncertainty of $\eta$ is thus expected to be at the percent level.
\end{itemize}

Some of these points cannot be addressed within the scope of this paper, in particular the angular distribution of the DPs and the specifics of the optics, as well as the modelling of the mirror imperfections. We thus postpone a full Monte Carlo simulation of the setup to future work; only such a complete simulation can lead to a precise forecast. The motivation for this paper is to present an outline of the idea and we thus content ourselves with this toy simulation.

In conclusion, we find that our assumptions in the main part are robust against our more detailed investigation. In particular, we find that our analytically estimated kinematic cutoffs are a solid choice, rendering the analytical results reliable.

\clearpage
\bibliographystyle{JHEP}
\bibliography{bib.bib}

\end{document}